\documentclass[a4paper,fleqn,usenatbib]{mnras}

\usepackage{mathptmx}

\usepackage[T1]{fontenc}
\usepackage{ae,aecompl}

\usepackage{graphicx}	
\usepackage{amsmath}	
\usepackage{amssymb}	

\title[New insights on NGC\,6720]{Optimal fitting of gaussian-apodized
  or under-resolved emission lines in Fourier Transform spectra
  providing new insights on the velocity structure of NGC\,6720}

\author[Thomas B. Martin et al.]{
Thomas B. Martin,$^{1}$\thanks{E-mail: thomas.martin.1@ulaval.ca}
Simon Prunet,$^{2}$
Laurent Drissen,$^{1}$
\\
$^{1}$Universit{\'e} Laval, 2325, rue de l'universit{\'e}, Qu{\'e}bec (Qu{\'e}bec), G1V 0A6, Canada\\
$^{2}$Canada-France-Hawaii Telescope, 65-1238 Mamalahoa Hwy, Kamuela,
Hawaii 96743, USA }

\date{Accepted XXX. Received YYY; in original form ZZZ}

\pubyear{2016}

\DeclareRobustCommand{\NII}{\textup{N\,\textsc{\lowercase{II}}}}
\DeclareRobustCommand{\SII}{\textup{S\,\textsc{\lowercase{II}}}}
\DeclareRobustCommand{\OIII}{\textup{O\,\textsc{\lowercase{III}}}}
\DeclareRobustCommand{\OII}{\textup{O\,\textsc{\lowercase{II}}}}
\DeclareRobustCommand{\kms}{km\,s$^{-1}$}
\begin{document}
\label{firstpage}
\pagerange{\pageref{firstpage}--\pageref{lastpage}}
\maketitle

\begin{abstract}
  An analysis of the kinematics of NGC\,6720 is performed on the
  commissioning data obtained with SITELLE, the Canada-France-Hawaii
  Telescope's new imaging Fourier transform spectrometer. In order to
  measure carefully the small broadening effect of a shell expansion
  on an unresolved emission line, we have determined a computationally
  robust implementation of the convolution of a Gaussian with a sinc
  instrumental line shape which avoids arithmetic overflows. This
  model can be used to measure line broadening of typically a few
  \kms{} even at low spectral resolution (R less than 5000). We have
  also designed the corresponding set of Gaussian apodizing functions
  that are now used by ORBS, the SITELLE's reduction pipeline. We have
  implemented this model in ORCS, a fitting engine for SITELLE's data,
  and used it to derive the [\SII] density map of the central part of
  the nebula. The study of the broadening of the [\NII] lines shows
  that the Main Ring and the Central Lobe are two different shells
  with different expansion velocities. We have also derived deep and
  spatially resolved velocity maps of the Halo in [\NII] and H$\alpha$
  and found that the brightest bubbles are originating from two
  bipolar structures with a velocity difference of more than
  35\,\kms{} lying at the poles of a possibly unique Halo shell
  expanding at a velocity of more than 15\,\kms{}.
\end{abstract}




\section{Introduction}
\label{sec:introduction}

The high quality of the data obtained during the recent commissioning
of SITELLE \citep{Drissen2010}, the Canada-France-Hawaii Telescope
(CFHT)'s new imaging Fourier transform spectrometer,
(\citet{Baril2016}, Drissen et al., in preparation) revealed that the
model of the emission lines used for the analysis of SITELLE's data
with its dedicated software, ORCS \citep{Martin2015}, was not
perfectly suitable because the original shape of the observed lines
was not taken into account in the model. Only the instrumental line
shape (ILS) was considered. We have thus designed a new line model
resulting from the convolution of a Gaussian line profile with the ILS
which could be used to measure very small broadening values (typically
a few \kms) even at low spectral resolution (R less than 5000). In
Section~\ref{sec:design}, we describe our new line model and determine
a more computationally robust implementation for small broadening
values which avoids arithmetic overflows. In Section~\ref{sec:gauss},
we provide the definition of a set of Gaussian apodizing functions
with an adjustable broadening parameter that is perfectly suitable for
this model. We then apply this new line model to a data cube of the
planetary nebula NGC\,6720 (the Ring nebula, M\,57), which was
SITELLE's first target during commissioning. This analysis provides
new insights on its velocity structure which are then presented in
section~\ref{sec:m57}.

\section{Design of a computationally robust model for a sinc line with Gaussian broadening}
\label{sec:design}

\subsection{The instrumental line shape of a Fourier transform
  spectrum}
\label{sec:ilssinc}

Although dispersive spectroscopy is, by far, the most widely used
technique to obtain spectra of astronomical targets, Fourier transform
spectrometers (FTS) have been successfully used, both on ground-based
telescopes and space observatories: SPIRE on Herschel
\citep{Griffin2010}, FIS-FTS on AKARI \citep{Kawada2008}, PFS on Mars
Express \citep{Formisano2005} and CIRS on the Cassini orbiter
\citep{Kunde1996}, BEAR at CFHT \citep{Maillard1982}, FTS-2 at the
James Clerk Maxwell Telescope \citep{Naylor2006}, all working at
infrared and sub-mm wavelengths. Our team has also developped SpIOMM,
SITELLE's prototype, at the Observatoire du Mont M{\'e}gantic
\citep{Grandmont2003, Bernier2006}. In all cases, the core of the
instrument is a Michelson interferometer, which acts exactly like an
inverse Fourier transform on the input light and the signal sampled at
its output is an interferogram. To recover the input signal's
spectrum, one has to apply a Fourier transform to the measured
interferogram. Combining the interferometer with imaging optics and a
detector array like a bolometer array or a charge-coupled device
(CCD), millions of spatially resolved interferograms of extended
sources can be recorded simultaneously by these imaging Fourier
transform spectrometers (iFTS). SpIOMM and SITELLE have pushed this
concept to the more technically challenging near-UV and visible
wavelengths (350 - 800 nm), wide fields of view ($11' \times 11'$),
and spaxel numbers (4 million for SITELLE).

\begin{figure}
  \includegraphics[width=\columnwidth]{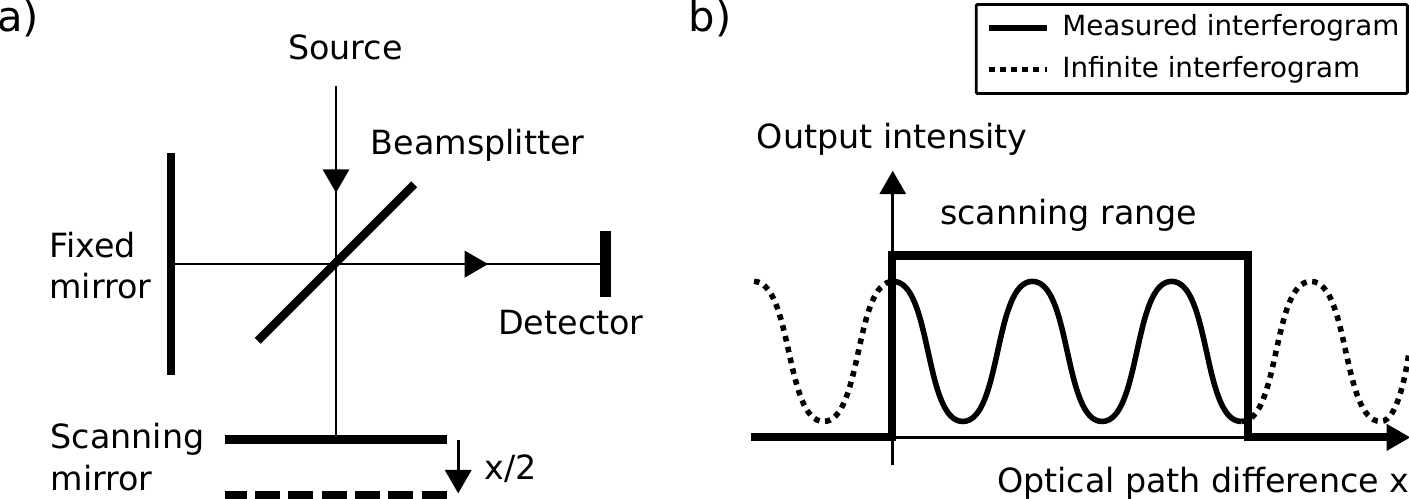}
  \caption{a) Sketch of a classical Michelson interferometer. When the
    scanning mirror is moved from a distance $x/2$ from the position
    where the two interfering beams have the same optical path, the
    OPD is equal to $x$. b) Ideal interferogram of a monochromatic
    source (a cosine). We also show the real scanning range
    represented as a rectangular function and the infinite
    interferogram which could be recorded if the scanning range was
    infinite.}
    \label{fig:michelson}
\end{figure}

Although SITELLE uses an off-axis configuration to allow the recording
of both output ports, the example of a classical Michelson
interferometer can be used to show how an interferogram is recorded
(see Fig.~\ref{fig:michelson}). The collimated beam at the input is
divided into two (ideally identical) parts by the beamsplitter. The
two beams are then reflected back onto the beamsplitter by two mirrors
(one is fixed while the other can be moved) where they interfere,
giving two complementary output beams: one is going back to the source
while the other can eventually be measured by a detector. During a
scan, the optical path difference (OPD, noted $x$ and measured in cm)
between the two interfering beams is gradually changed, step by step,
modulating the output intensity according to the spectral energy
distribution of the observed source. The intensity recorded at each
step forms the interferogram; its Fourier transform must be computed
to recover the spectrum of the source.

One of the most distinctive features of a spectrum obtained with an
FTS is the ILS which is a sinc\footnote{Throughout this paper the sinc
  function is defined as $\text{sinc}(\sigma)=\frac{\sin{(\pi
      \sigma)}}{\pi \sigma}$.}. It
comes from the fact that the number of observed samples is finite thus
implying a finite maximum value of the OPD. Any measured interferogram
$I(x)$ can thus be considered as an infinite interferogram
$I_\infty(x)$ multiplied with a rectangular function $\Pi(x)$:
\begin{equation}
  I(x) = I_\infty(x)\Pi(x)
\end{equation}
As the spectrum is the Fourier transform of the interferogram, the
convolution theorem states that the spectrum will be the convolution
of the real spectrum at an infinite resolution with the Fourier
transform of a rectangular function, i.e. a pure sinc function.
\begin{alignat}{2}
  S(\sigma)&=TF[I(x)] = TF[I_\infty(x)] * TF[\Pi(x)]\\
  &=S_{\text{R}=\infty}(\sigma) * \text{sinc}(\sigma)\;.
\end{alignat}
If the spectrum is a monochromatic line of flux unity and wavenumber
$\sigma_0$, its spectrum is similar to a Dirac delta-function
$\delta(\sigma - \sigma_0)$ and the resulting measured spectral line
$S_{\sigma_0}(\sigma)$ is, by definition, the instrumental line shape.
\begin{equation}
  S_{\sigma_0}(\sigma)=\delta(\sigma - \sigma_0) * \text{sinc}(\sigma) = \text{sinc}(\sigma - \sigma_0)
\end{equation}

The ILS of an ideal FTS can fully be described via its full-width at
half-maximum (FWHM) which limits the resolution of the convoluted
spectrum and depends only on the maximum OPD (MPD) reached during the
interferogram recording. The equation giving the width of the sinc,
$\Delta w$ (in cm$^{-1}$), from the MPD of the interferogram (in cm),
can be written as \citep{Kauppinen2001}
\begin{equation}
  \label{eq:delta_w}
  \Delta w = \frac{1}{2\text{MPD}}\;,
\end{equation}
with the FWHM of the sinc function, $\text{FWHM} = 1.20671\;\Delta w$ (see equation~\ref{eq:sinc}).
At a given wavenumber $\sigma$ (in cm$^{-1}$), the resolution $R$
is therefore \citep{Martin2015-thesis}:
\begin{equation}
  \label{eq:resolution}
  R = \frac{\sigma}{\text{FWHM}} = \frac{\sigma}{1.20671\;\Delta w} = \frac{2\sigma\;\text{MPD}}{1.20671}\,.
\end{equation}
On the optical axis of the interferometer, the MPD is the
product of the optical step size, $\delta_x$, by the number of
steps, $N$, on the largest side of the interferogram with respect to
the zero path difference. But if we are recording an interferogram at
a certain angle $\theta$ a cosine term must be taken into account.
\begin{equation}
  \text{MPD} = N\delta_x\cos{\theta}
\end{equation}

The knowledge of the ILS is very important if we want to fit emission
and absorption lines in a Fourier transform spectrum
\citep{Martin2015}. Even if we know that the model of a pure
monochromatic line is a sinc (with an ideal instrument), perfectly
monochromatic lines do not exist. Their shape is mostly determined by
their emission mechanism (internal motions in an HII region, for
instance). Therefore, the observed line shape is always the
convolution of the ILS with the intrinsic line shape as it ends up on
our detectors.

\subsection{General model of a sinc ILS with Gaussian broadening}
\label{sec:ils}

Due to thermal Doppler broadening, star rotation or gas motion
broadening, the vast majority of the observed lines, in emission as
well as in absorption, have a Gaussian shape. At low resolutions (R
$<$ 1000), the width of the line is generally small enough compared
with the width of the ILS that a pure sinc function can be considered
as a good model for line fitting. But, as we will see in this section,
at higher resolution or when the line broadening is sufficiently high,
a line will start to exhibit a larger profile which is better
modelized as the convolution of a sinc, $S_{\Delta w}(\sigma -
\sigma_0)$, of width $\Delta w$ centered on the wavenumber of the line
($\sigma_0$) and a Gaussian, $G_{\Delta \sigma}(\sigma)$, of width
$\Delta \sigma$,
\begin{equation}
  \label{eq:convolve}
  SG(\sigma) = S_{\Delta w}(\sigma) * \delta(\sigma - \sigma_0)* G_{\Delta \sigma}(\sigma)\;,
\end{equation}
with,
\begin{equation}
\label{eq:sinc}
  S_{\Delta w}(\sigma) = \text{sinc}(\frac{\sigma}{\pi\Delta w}) = \frac{\sin(\sigma/\Delta w)}{\sigma/\Delta w}\;,
\end{equation}
and,
\begin{equation}
  G_{\Delta \sigma}(\sigma) = \exp(\frac{-\sigma^2}{2\Delta\sigma^2})\;.
\end{equation}

The two key parameters of this model are: the width, $\Delta w$, of
$S_{\Delta w}(\sigma)$, which only depends on the resolution (see
equation~\ref{eq:resolution}), and the line broadening which gives the
width of $G_{\Delta \sigma}(\sigma)$. It is sometimes better to
consider the broadening ratio $\Delta\sigma / \Delta w$ because it
provides a measurement of the line broadening which is independent of the
resolution.

An analytic formulation of this convolution is well-known and is
already implemented in the Herschel Interactive Processing Environment
\citep{Ott2010}. But we will show in the next section that its
implementation is not optimal because of an overflow error that limits
its use to large broadening ratios. We can start by giving this
formulation as it is written in the Herschel Common Science System
(HCSS)
documentation\footnote{\url{http://herschel.esac.esa.int/hcss-doc-14.0}}.
\begin{equation}
  \label{eq:erf_form}
  SG(\sigma)_{\text{erf}} = A\,e^{-b^2}\,\frac{\text{erf}(a - ib) + \text{erf}(a + ib)}{2\,\text{erf}(a)}\;,
\end{equation}
with 
\begin{equation}
  \label{eq:a_b}
  a = \frac{\Delta\sigma}{\sqrt{2} \Delta w}\,\text{ and }b = \frac{\sigma - \sigma_0}{\sqrt{2} \Delta\sigma}\;,
\end{equation}
where $A$ is the maximum value of the function, $\sigma_0$ its maximum
position, $\Delta w$ is the width of the sinc and $\Delta \sigma$ the
width of the Gaussian. This form will be considered as the basic erf
formulation of the gaussian-sinc convolution (a demonstration of this
result is given in Appendix~\ref{appendix_demo}).

\begin{figure}
  \includegraphics[width=\columnwidth]{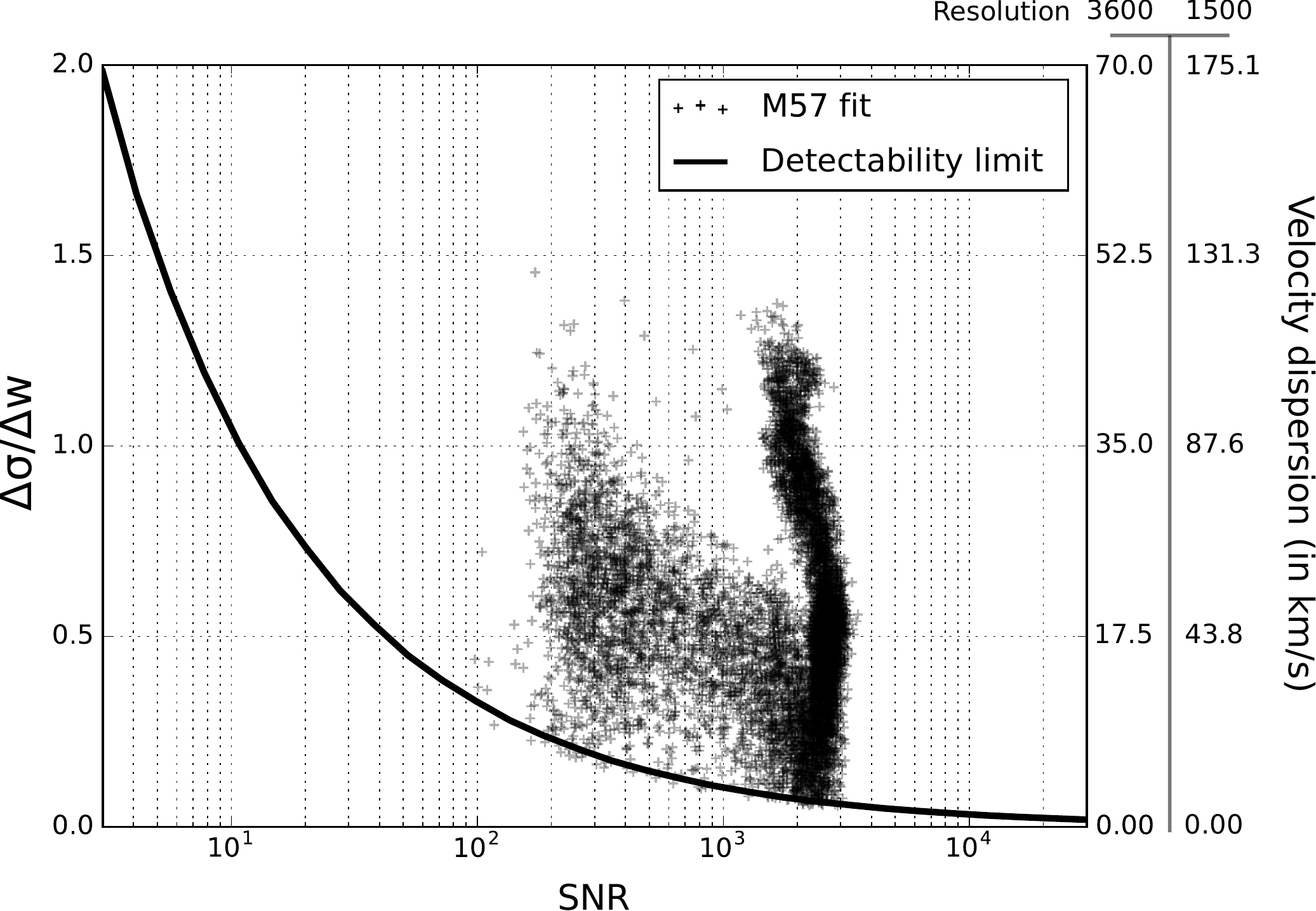}
  \caption{Minimum measurable broadening ratio $\Delta\sigma/\Delta w$
    for a line with a given SNR. The corresponding velocity dispersion
    around H$\alpha$ (656.28\,nm) at two different resolutions (3700
    and 1500) has been reported on the figure. It scales
    linearly with the resolution. A randomly chosen set of points from
    the NGC\,6720 cube fit is shown. The SNR of the data points is high
    because all the lines in the spectrum are fitted with the same
    broadening parameter.}
    \label{fig:ratiovssnr}
\end{figure}

As mentioned above, if the line broadening is sufficiently small a
pure sinc can be used as a reliable model. Then, at a given resolution
there must exists a limit in signal-to-noise ratio (SNR) where the
broadening is not measurable any more, i.e. when the difference, point
by point, between a broadened line and a pure sinc becomes smaller
than the noise standard deviation. This detectability limit is shown
in Fig.~\ref{fig:ratiovssnr} in terms of broadening ratio.
The velocity dispersion $\Delta v$ corresponding to this broadening
ratio has also been computed using the following equation:
\begin{equation}
  \label{eq:velocity}
  \Delta v = c\frac{\Delta\sigma}{\sigma}
\end{equation}
It clearly appears that, when the resolution decreases, the minimum
broadening ratio needed to measure the same velocity dispersion
becomes smaller. At a resolution of 1500, a bubble expanding at a few
\kms{} will broaden a line with a SNR of 1000. Therefore, even at small
resolution, we must be able to model an emission line with a small
broadening ratio. In the next section we will show that the
formulation given in equation~\ref{eq:erf_form} is not suitable
for modelling small broadening ratios and we will propose a better
formulation in this respect.

\subsection{A more computationaly robust analytic formulation}
\label{sec:daws}
Starting from equation~\ref{eq:erf_form}, we can derive another
formulation of $SG(\sigma)$ by using the definition of a Dawson integral
$D(\sigma)$:

\begin{equation}
  \label{eq:dawson_def}
  D(\sigma) = e^{-\sigma^2} \int_0^\sigma e^{y^2} dy = \frac{-i\sqrt{\pi}}{2}\,e^{-\sigma^2} \text{erf}(i\sigma)\,.
\end{equation}

Replacing the error function by a Dawson integral everywhere in
equation~\ref{eq:erf_form} gives what we will call the Dawson
formulation of $SG(\sigma)$:

\begin{equation}
  \label{eq:daws_form}
  SG(\sigma)_{\text{Dawson}} = A  \frac{D(ia + b)\,e^{2iab} + D(ia - b)\,e^{-2iab}}{2D(ia)}\,.
\end{equation}

\begin{figure}
  \includegraphics[width=\columnwidth]{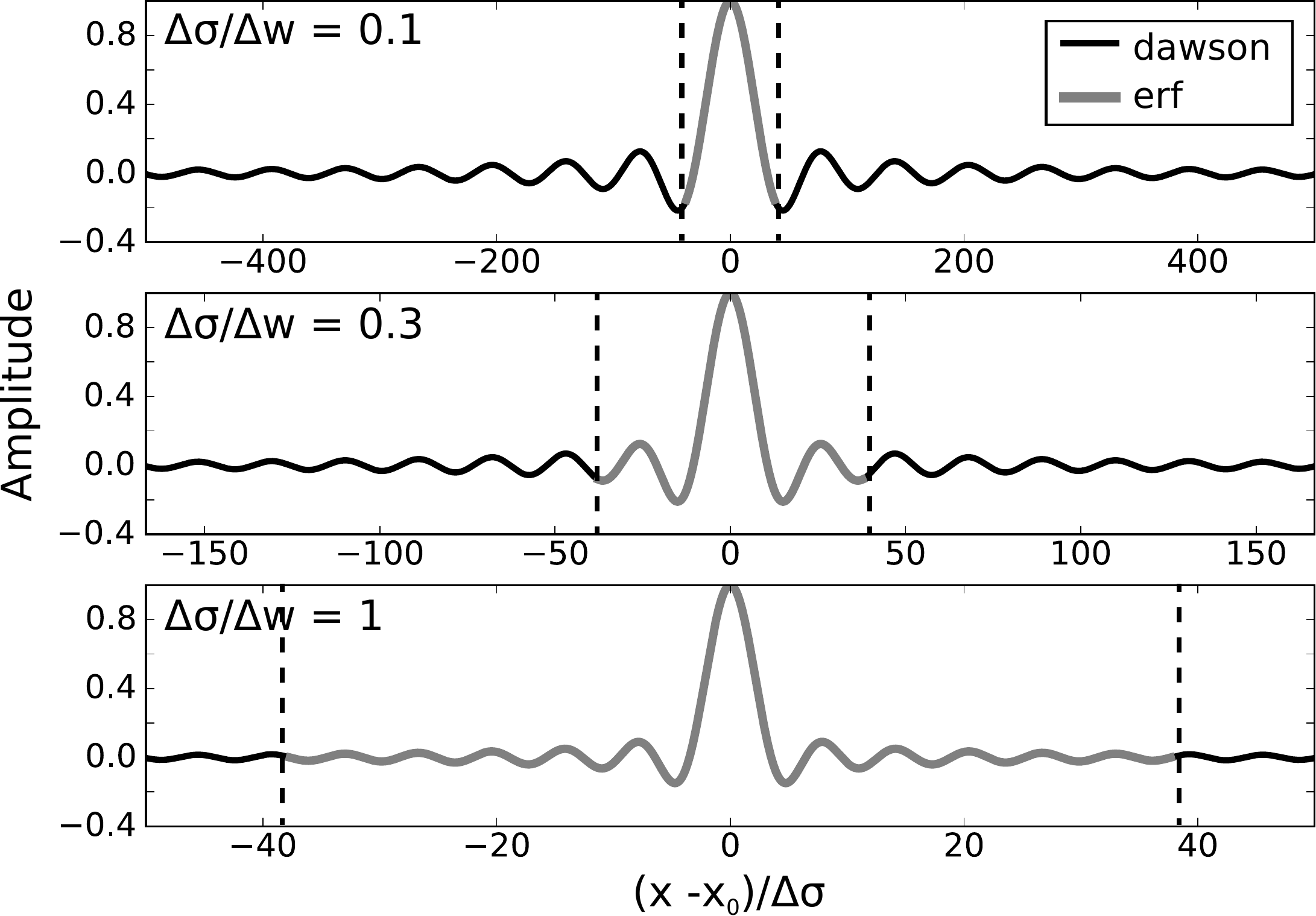}
  \caption{Comparison of Dawson and erf models for different values of
    $\Delta\sigma/\Delta w$. The truncation of the erf model due to
    overflow errors when $b > 26.64$ (equivalent to
    $(\sigma-\sigma_0)/\Delta\sigma > 37.67$) appears clearly.}
    \label{fig:example_dsigma}
\end{figure}

\begin{figure}
  \includegraphics[width=\columnwidth]{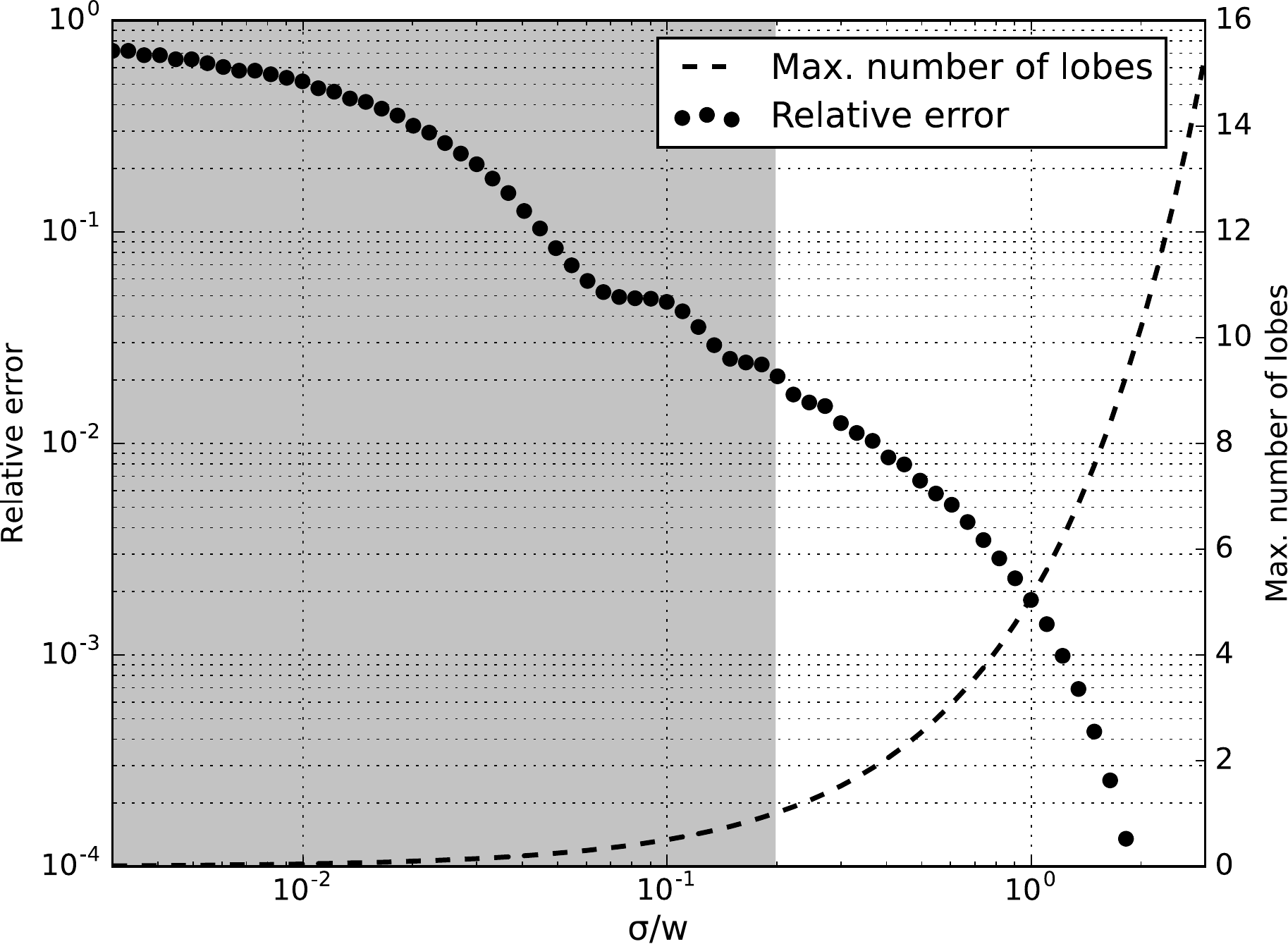}
  \caption{Limitations of the erf model. Due to overflow errors
    $SG_{\text{erf}}(\sigma)$ cannot be computed too far from the central
    lobe. The maximum number of lobes on one side of the function that
    can be represented and the relative error made because of the
    truncation of the model are presented (see text for details on the
    error evaluation). The region where even the central lobe is not
    fully evaluated is displayed in grey.}
    \label{fig:erf_vs_dsigma}
\end{figure}

Overflow arises for both the error function and the Dawson integral
when the imaginary part of the input becomes greater than 26.64 in
64-bit float representation (the limit is around 9.4 for 32-bit
floats)\footnote{$\text{erf}(26.64i) > 3.4\,10^{306}i$ and $D(26.64i)
  > 1.4\,10^{308}i$. The 64-bit floating limit in python is around
  $1.7\,10^{308}$.}. The erf model is thus limited by the value of $b$
which scales as $(\sigma-\sigma_0)/\Delta\sigma$. When $\Delta\sigma$ is small
compared to the width of the sinc, the erf model cannot be evaluated
after a few lobes (typically less than 1 if $\Delta\sigma < 0.27
\Delta w$, see Fig.~\ref{fig:example_dsigma}) where their relative
amplitude is significantly high.
When fitting a line, the truncated part can only be replaced with
zeros as it is the case in the HCSS. It is possible to evaluate the
relative error $\epsilon$ as the ratio of the energy contained in the
non-truncated part over the total energy of the line:
\begin{equation}
  \label{eq:relerror}
  \epsilon = \frac{\int_{-\infty}^{+\infty}{SG_{\text{erf}}(\sigma)^2}\,\text{d}\sigma}{\int_{-\infty}^{+\infty}{SG_{\text{Dawson}}(\sigma)^2\,\text{d}\sigma}}
\end{equation}
We can see in Fig.~\ref{fig:erf_vs_dsigma} that this error can be
greater than 70\% for small values of $\Delta\sigma/\Delta w$ because
when ($\Delta\sigma < 0.2 \Delta w$), even the central lobe cannot be
fully evaluated (see Fig.~\ref{fig:example_dsigma}). 

Note that the Dawson model is also subject to an overflow error when
$\Delta\sigma/ \Delta w > 37$. Such a high broadening level
corresponds to an expansion velocity greater than 3500\,\kms{} at a
resolution of 1500, which is easily observable in the AGN of numerous
galaxies \citep{OsterbrockDonaldE.2006}. It could be fitted with a
simple Gaussian model but it's interesting to note that in this case
the erf model can be used instead of the Dawson model.

\section{Design of a Gaussian apodizing functions for Fourier Transform Spectroscopy}
\label{sec:gauss}
Apodization is frequently used to reduce the amplitude of the
sidelobes of the ILS to obtain a more dispersive-like spectrum. The
idea is to convolve the ILS with a function designed to have the
desired properties. A convolution in the spectrum space is a
product in the interferogram space and, most of the time, this
operation is better done on the interferogram before the Fourier
transform (see, for example, \citealt{Davis2001}). $A(x)$ being the
apodizing function and $I(x)$ the original interferogram, the
apodization operation can be simply written as
\begin{equation}
  \label{eq:apodization_general}
  I_{A}(x) = I(x)A(x)\;.
\end{equation}

We have seen in the previous sections that the convolution of a sinc
and a Gaussian will broaden the ILS and reduce the size of the
sidelobes. Broadening the ILS will result in a reduced spectral
resolution while the general shape of the instrumental line will
become more Gaussian. If the ILS is well known (a sinc in the case of
an ideal FTS) there is absolutely no need to apodize the
spectrum. But, if, for some reasons, the ILS cannot be perfectly
modeled, as it happens with non-ideal instruments, apodizing will
smooth the ILS and help in reducing fitting artifacts coming from a
bad model.

A lot of apodizing functions can be found in the literature (e.g. see
\citealt{Naylor2007} and references therein), each one of them having
its own particular use and properties. But when it comes to fitting,
having the best model possible is of great importance. If we consider
an ideal instrument with an ideal ILS, the ILS of an apodized spectrum
can be described as being the result of the convolution of a sinc and
the Fourier transform of the apodizing function. An analytic
formulation of this ILS can be written as
\begin{equation}
  \label{eq:base_model}
  M(\sigma) = S_{\Delta w}(\sigma)*FT[A(x)](\sigma)\;.
\end{equation}
Now, if we want to analyze a Gaussian-shaped line of width
$\Delta\sigma_1$ (see previous section) a more realistic model would
be
\begin{equation}
  \label{eq:real_model}
  M(\sigma) = G_{\Delta \sigma_1}(\sigma)*S_{\Delta w}(\sigma)*FT[A(x)](\sigma)\;.
\end{equation}

Having an analytic formulation of such a fitting model can be
difficult or even impossible to derive depending on the complexity of
$FT[A(x)](\sigma)$ . But if we consider using Gaussian apodizing
functions of width ${\Delta \sigma_2}$ (and remembering that the
Fourier transform of a Gaussian is also a Gaussian), the model of the
apodized spectrum becomes
\begin{eqnarray}
  M(\sigma)&=&G_{\Delta \sigma_1}(\sigma)*S(\sigma)*FT[G(x)](\sigma)\\
  &=&G_{\Delta \sigma_1}(\sigma)*S(\sigma)*G_{\Delta \sigma_2}(\sigma)\\
  M(\sigma)&=&G_{\Delta \sigma_{1*2}}(\sigma)*S(\sigma)\;,
  \label{eq:real_model_gauss}
\end{eqnarray}
where,
\begin{equation}
  \Delta \sigma_{1*2} = \sqrt{\Delta \sigma_1^2 + \Delta \sigma_2^2}\;.
  \label{eq:widthconv}
\end{equation}
Equation~\ref{eq:real_model_gauss} comes directly from the fact that
the convolution of two Gaussians is a Gaussian with a width given by
equation~\ref{eq:widthconv}. The fitting model developed in the
previous section can therefore also be used  on Gaussian-apodized
spectra. Even if Gaussian apodized functions may not have the best
properties in terms of sidelobes magnitude reduction for a given
broadening \citep{Harris1978} they are still interesting in this
respect. Following the idea introduced by \cite{Naylor2007}, we have
designed Gaussian apodizing functions with an adjustable broadening
factor $b$, i.e. the factor by which the FWHM is increased when the
spectrum is apodized. They can be described as simple Gaussian
function,
\begin{equation}
  \label{eq:gaussian_apod}
  G(w, x) = \exp{\left(-\frac{x^2}{\text{MPD}^2}w(b)^2\right)}\;,
\end{equation}
where,
\begin{equation}
  \label{eq:b_factor}
  w(b) = b - 1 + 2\sqrt{2\ln 2}\;\text{erf}\left(\frac{\pi \sqrt{b - 1}}{2}\right)\;.
\end{equation}

Because the resulting ILS of the spectra apodized with this set of
functions can be perfectly modeled with the model derived in
Section~\ref{sec:daws} (see equation~\ref{eq:daws_form}) they have
been implemented in ORBS, SITELLE's reduction pipeline
\citep{Martin2012}, as the default apodizing function.

\section{New insights on the velocity structure of NGC\,6720}
\label{sec:m57}

\defcitealias{Odell2013a}{Od13}
\defcitealias{Odell2007}{Od07}
\defcitealias{Guerrero1997}{Gu97}

\begin{figure*}
  \includegraphics[width=\linewidth]{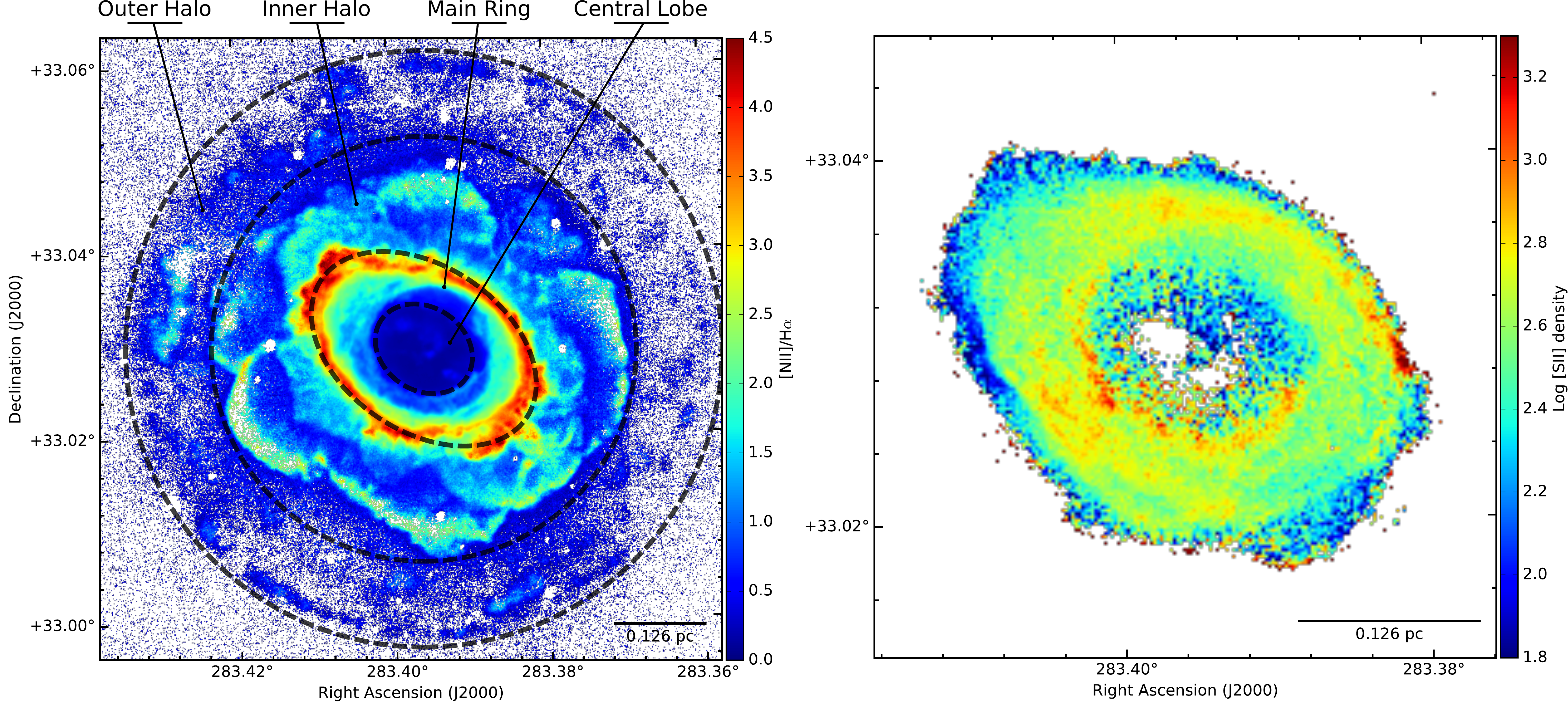}
  \caption{\textit{Left}: Map of the  [\NII]$\lambda$6584 /
    H$\alpha$ line ratio of NGC\,6720. The approximate contours of the different
    regions of the nebula are indicated. \textit{Right}: Map of the
    [\SII] density of the central part of NGC\,6720. Following
    \citet{Odell2013c}, this map has been computed with the relation
    $\log n_e = 4.705 - 1.9875\,S(671.6) / S(673.1)$ from
    \citet{OsterbrockDonaldE.2006} which is accurate for densities in
    the range 100\,cm$^{-3}$ -- 3000\,cm$^{-3}$. The uncertainty on
    the estimate of the [\NII]/H$\alpha$ ratio and the density is
    shown in Fig.~\ref{fig:ratios-err}.}
  \label{fig:ratios}
\end{figure*}

\begin{figure}
  \includegraphics[width=\linewidth]{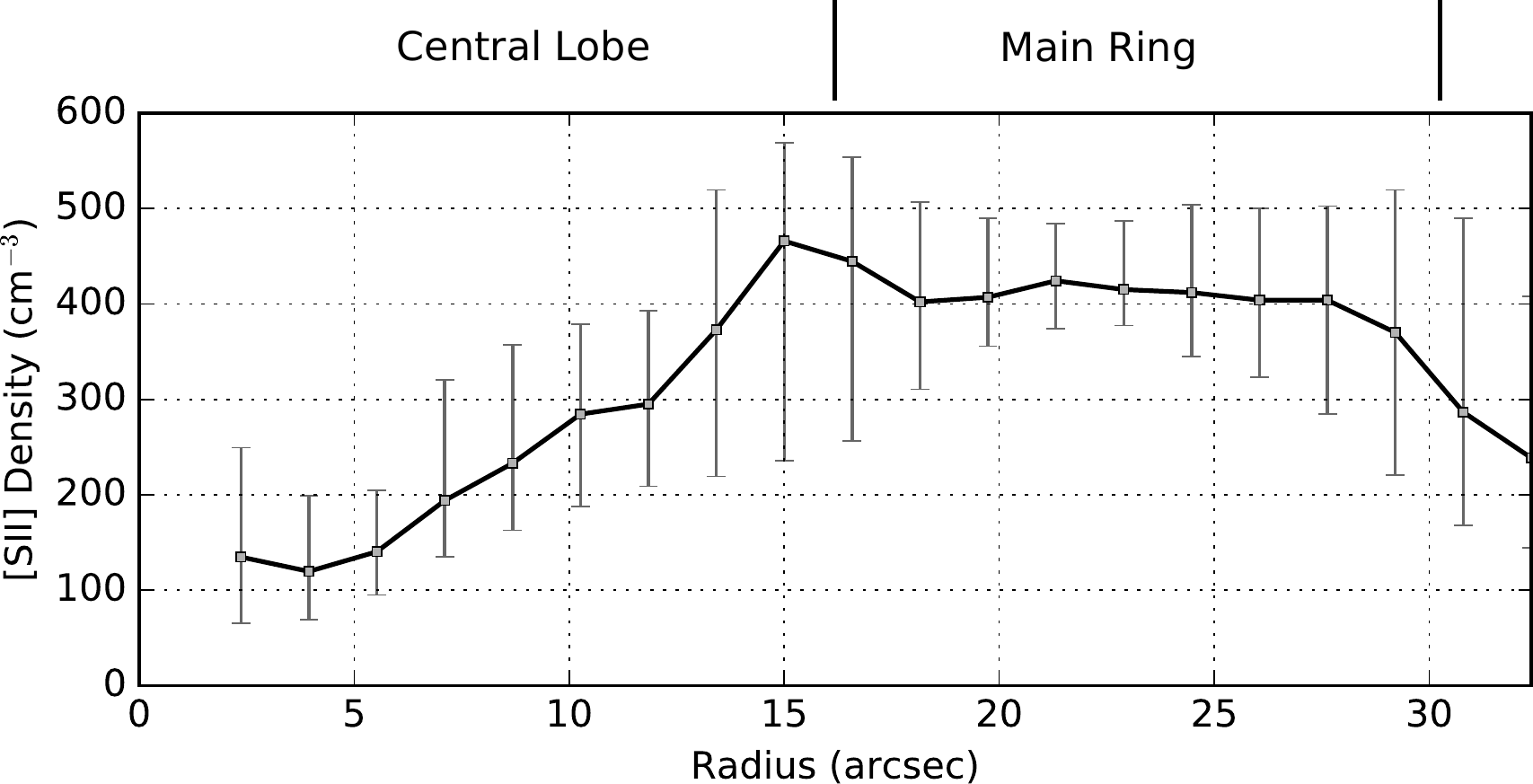}
  \caption{Elliptical profile of the [\SII] density of the central
    part of NGC\,6720 considering a mean temperature of
    10\,000\,K. The density map has been convoluted with a $3\times3$
    Gaussian kernel. Density has been computed from the [\SII] ratio
    with Pyneb \citep{Luridiana2012}, the Python implementation of the
    nebular package \citep{Shaw1995}. The radius is given for the
    minor axis of the ellipses. The parameters of the ellipse scaled
    to obtain this profile is the Main Ring ellipse shown on
    Fig.~\ref{fig:ratios}. The position of the Central Lobe ellipse
    and the Main Ring ellipse along the minor axis are indicated. The
    error bars show the 16th and 84th percentiles of the distribution
    along the ellipses.}
  \label{fig:density}
\end{figure}

NGC\,6720 (also known as M\,57 or the Ring nebula) is one of the
brightest planetary nebulae (PNe) of the Galaxy. It has been
extensively studied by numerous authors and techniques, making it an
object of choice to evaluate the quality of the results obtained with
the model depicted in the previous sections and try to provide new
insights on its general structure. We refer especially to the most
recent and extensive study of (\citealt{Odell2013a}, henceforth
\citetalias{Odell2013a}) for a complete description of this object.
As a general picture, this nebula is seen along its polar axis, so
that its equatorial plane is nearly parallel to the plane of the
sky. It consists of a dim core made of two lobes expanding along the
polar axis and surrounded by a bright lower ionization region of
higher density called the Main Ring which lies in the equatorial plane
of the object. Both regions are optically thick and ionization bounded
\citepalias{Odell2013a}. All around the Main Ring we find the Inner Halo,
recognizable at its petal-like structures which may be thin bubbles of
ionized material (\citealt{Guerrero1997}, henceforth
\citetalias{Guerrero1997}, and \citetalias{Odell2013a}).  The Outer
Halo is the farthest detectable structure. It presents a circular
shape on the plane of the sky which suggests that it is a thin
shell. Both the Inner and the Outer Halo features are considered to be
fossil radiation by \citetalias{Odell2013a}. The location of the
different regions is shown in Fig.~\ref{fig:ratios}. If we examine the
elliptical profile of the density shown in Fig.~\ref{fig:ratios}, the
Main Ring corresponds to the region where the density is at its
highest level compared to the Central Lobe and the Halo (see
Fig.\ref{fig:density}).


\subsection{Observations with SITELLE}

NGC\,6720 has been observed with SITELLE on August 8, 2015 during the
commissioning of the instrument at CFHT (L. Drissen et al. 2016, in
preparation, \citealt{Baril2016}). A data cube has been obtained in
the SN3 filter (647 - 685 nm, including H$\alpha$ as well as the [NII]
and [SII] doublets) with a resolution R = 2600 and a seeing-limited
spatial resolution of 0.8'' (see Table~\ref{tab:obsparams}). This data
has been fully reduced and calibrated with ORBS (Data Release 1, see
T. Martin et al. 2016, in preparation and \citealt{Martin2012}).  Flux
calibration and transmission correction have been made from the
observation of the spectrophometric standard planetary nebula M1-71
\citep{Wright2005} obtained during the same observing run, as well as
the observation of the spectrophotometric standard star GD71
\citep{Bohlin2001} obtained in January 2016 through the same filter.
A primary wavelength calibration has been done via the observation of
a green He-Ne laser source at 543.5\,nm. This calibration has been
refined by the measurement of the sky spectra, more specifically using
the numerous Meinel OH bands present in the filter passband, around
NGC\,6720 and corrected for heliocentric velocity with ORCS (Martin et
al. 2016, in preparation). The pixel-to-pixel calibration error on the
velocity is estimated to be below 0.5\,\kms. The absolute velocity
calibration is estimated to be around 1\,\kms{} (see
appendix~\ref{sec:abs_vel_calib} for details).
\begin{table}
  \caption{Observation parameters of NGC\,6720. Resolution is computed from the number of steps with respect to the ZPD position ($593 - 215 = 378$) at the H$\alpha$ wavelength and for an off-axis angle of 15.4$^{\circ}$ that corresponds to the center of the field of view.}
 \label{tab:obsparams}
 \begin{tabular}{lccccc}
  \hline
  Filter & Order & Step & Step & ZPD & R @H$\alpha$\\
  &&size (nm)&number&position&($\theta$=15.4$^{\circ}$)\\
  \hline
  SN3 & 8 & 2880.88&593&215&2650\\
  \hline
 \end{tabular}
\end{table}

\begin{figure}
  \includegraphics[width=\columnwidth]{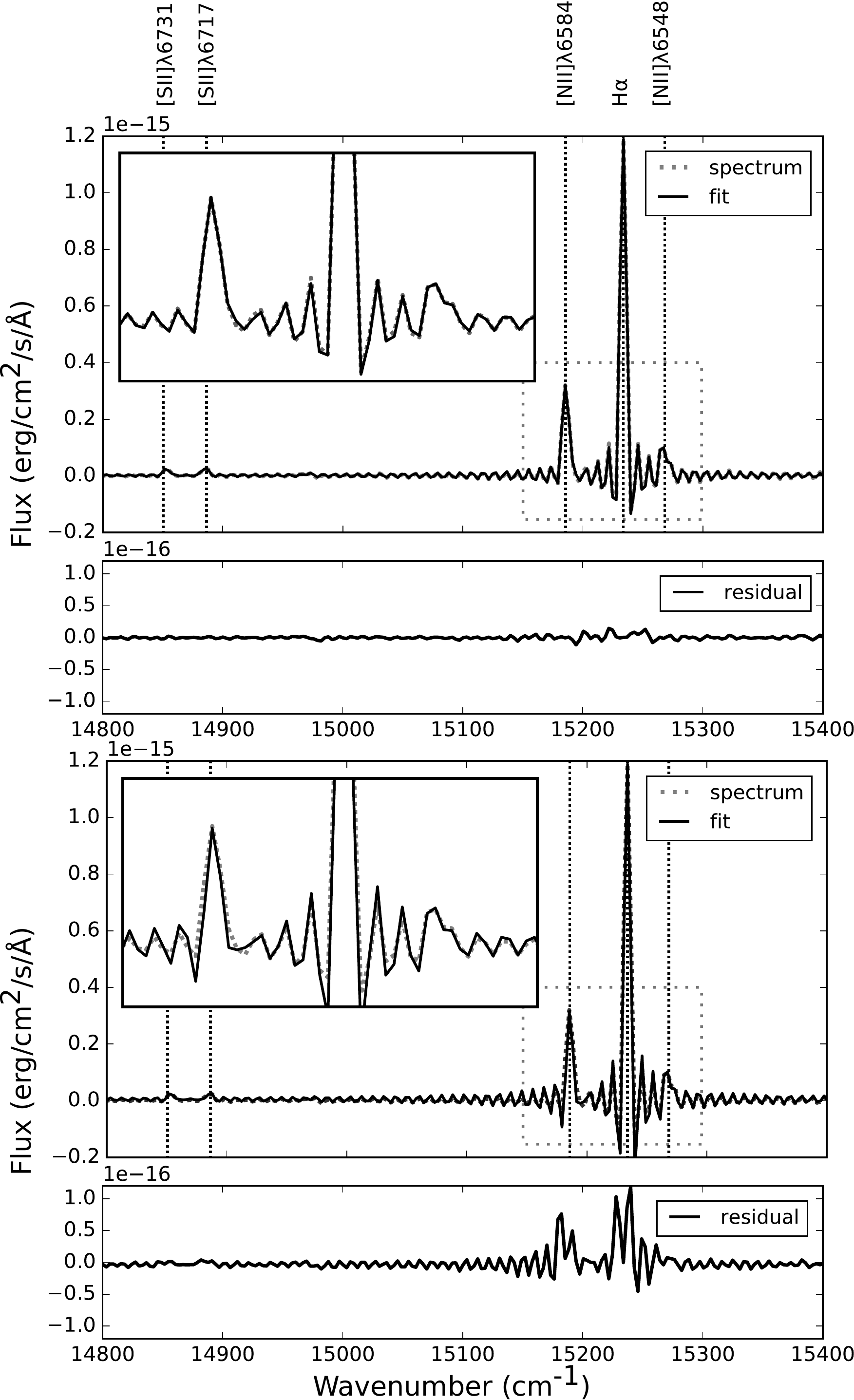}
  \caption{\textit{Top}: Example of a fit on one NGC\,6720 spectrum
    realized with the $SG_{\text{Dawson}}(\sigma)$ model developed in
    this article. The SNR of the lines ranges from 600, for H$\alpha$,
    to 20 for the [SII] lines. The fit, superimposed on the original
    data, and the residuals, are shown. A zoom on the H$\alpha$ region
    illustrates the quality of the model used. \textit{Bottom}: Fit of
    the same spectrum as on the top panel but realized with a pure
    sinc model. The high residual near the first lobes of the lines
    indicates a poorly fitted model.}
    \label{fig:m57_fit}
\end{figure}

With an integration time of 2.5 hours, the SNR of the spectra covering
most of the nebula is high enough to study the dynamics with some
precision: individual spectra in the brightest regions exhibit
lines of H$\alpha$ and [\NII]$\lambda$6583 with a SNR greater than
100. One of the approximately 280\,000 spectra, taken near the center
of nebula (where the velocity dispersion is the highest), is shown in
Fig.~\ref{fig:m57_fit}.

\subsection{Evaluation of the fitting model}

To illustrate the importance of using a model based on the convolution
of a sinc and a Gaussian we show in Fig.~\ref{fig:m57_fit} the
results of two fits made on this spectrum. One of the fits makes use
of the line model developed in the previous section and the other is
based on a pure sinc. The residual near the first lobes of the
emission line is more than ten times higher when using a pure sinc
model and exhibits a typical modelling error. The implementation of
the Dawson model in ORCS, the fitting engine developed for SITELLE's
data \citep{Martin2015}, is robust enough to be used on a regular
basis to fit entire 3D cubes. With this software we have mapped the
velocity dispersion of the ionized gas in NGC\,6720. Because the
angular size of a pixel is 0.32\,'', which is more than two times
smaller than the average seeing of the observation, the spectral
cube has been binned 2x2 to enhance the SNR without losing too much
spatial resolution.

\begin{figure*}
  \includegraphics[width=\linewidth]{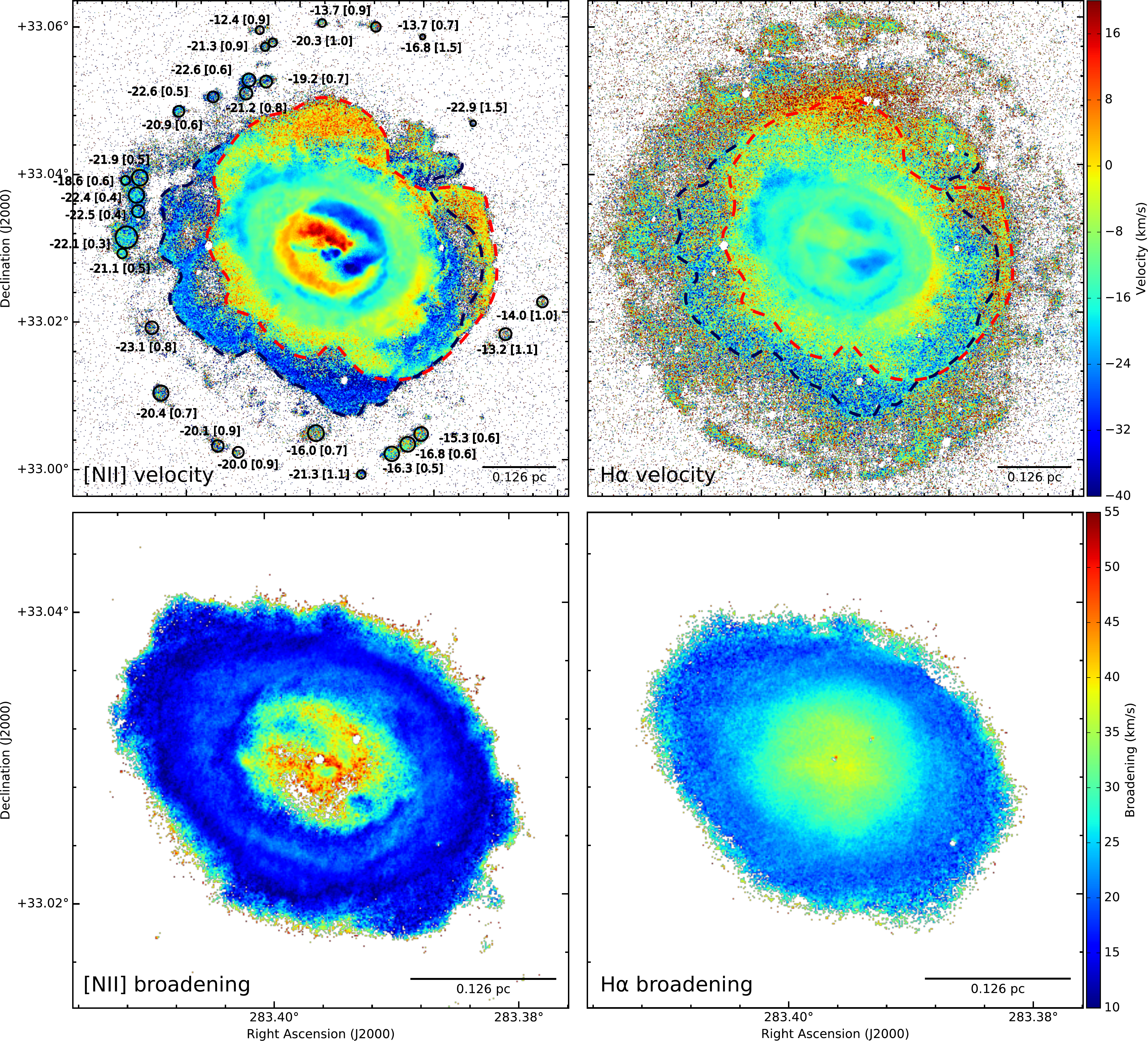}
  \caption{Velocity (top) and broadening (bottom) maps
    of[\NII]$\lambda$6584 and [\NII]$\lambda$6548 (left) and H$\alpha$
    (right). The two [\NII] lines have been fitted together with the
    same velocity and broadening parameters to enhance the precision
    of the fit. Color scales are indicated on the right of the
    figure. Maps of the uncertainties on these estimates of the
    velocity and the broadening are shown in
    Fig.~\ref{fig:maps-err}. \textit{Top-left}: The velocity of the
    ionized features at the boundary of the Outer Halo has been
    computed from the integrated spectra. The aperture used for each
    region is indicated by a black circle along with the estimate of
    the velocity and its uncertainty. The approximate boundaries of
    the Halo bipolar structures moving away and towards us are
    indicated with dotted lines respectively in red and blue. Those
    boundaries are reported in the H$\alpha$ velocity map (top-right
    panel).}
  \label{fig:maps}
\end{figure*}
As it is shown in Fig.~\ref{fig:maps} and Fig.~\ref{fig:maps-err},
the velocity dispersion along the line of sight has been measured on
all the spectra covering the object with an uncertainty as low as a
few \kms{}. Modelling the spectra over the entire spectral range covered by the filter is key for this
fitting procedure because the secondary lobes of individual lines are
interfering with each other up to the filter edges. It is thus important
to model a line with all its lobes even for small broadening
ratios. In the case of NGC\,6720, the emission lines with a velocity
dispersion around 10\,\kms{} have a broadening ratio of 0.2. With the
erf implementation even the first lobe could not have been computed
(see Fig.~\ref{fig:example_dsigma}).

ORCS uses a least-square Levenberg-Marquardt minimization algorithm
\citep{Levenberg1944,Marquardt1963} to fit the data. All the lines are
fitted simultaneously and some parameters are grouped together to
enhance the quality of the fit. The velocity and the broadening are
considered to be the same for the [\NII] and [\SII] lines (which
reduces the number of velocity parameters from 4 to 1). H$\alpha$
velocity and broadening are fitted independently from the other
lines. The Levenberg-Marquardt algorithm assumes the Gaussianity of
the noise distribution and the fact that the contours of the objective
function are ``nearly elliptical in the immediate vicinity of the
minimum'' \citep{Marquardt1963}. It is important to test the validity
of this hypothesis if we want to make sure that the estimation on the
parameters is unbiased and that the calculated uncertainty (derived
from the covariance matrix) is correctly estimated. We have used a
Monte-Carlo-Markov-Chain algorithm implemented in the emcee Python
library \citep{Foreman-Mackey2012} to get the marginalized
distribution of the posterior probability of each parameter (see
Fig.~\ref{fig:triangle}). We find that the amplitude and velocity
parameters show a circular 2D distribution around the best value and a
nearly Gaussian marginalized distribution. When the value of the
broadening parameter is sufficiently high (e.g. the [\NII], [\SII]
broadening which is around 12\,\kms{} in the example on the figure)
the distribution is nearly ideal. But it shows an asymmetric
distribution when it is approaching zero (which is the case of the
H$\alpha$ line). It means that very low broadening values (under a few
\kms{} down to zero) can be systematically biased to slightly higher
values.

\subsection{Study of the velocity dispersion in NGC\,6720}

\subsubsection{Limitations}
\label{sec:limitations}

Before we start any analysis of the broadening and velocity maps
obtained with SITELLE (see Fig.~\ref{fig:maps} and
Fig.~\ref{fig:maps-err}), we must consider some inherent limitations
to the use of low-resolution data. At a resolution of 2600, in the
case of a PNe like NGC\,6720, line broadening has 3 main physical
origins, all of the same order of magnitude: shell expansion (a few
10\,\kms{}), thermal broadening\footnote{The broadening values are
  expressed in terms of Gaussian width. In terms of FWHM we have
  21.5\,\kms{} for H$\alpha$ and 5.7\,\kms{} for [\NII]$\lambda$6584
  at a temperature of 10\,000\,K.} (9.1\,\kms{} for H$\alpha$ and
2.4\,\kms{} for [\NII]$\lambda$6584 at a temperature of 10\,000\,K)
and possibly turbulence (also around 10\,\kms{} see
\citealt{Odell2013c, Odell1991, medina2014} and references
therein). Another source of line broadening is related to the
instrument and comes from the modulation efficiency loss at large
OPD. This problem is well documented in \citet{Baril2016} and might
account for the larger measured velocity expansion of the shells (see
section~\ref{sec:elliptical_prof}). The commissioning data has been
taken without tilt correction so that we can expect a modulation
efficiency loss of up to 25\,\% at the MPD (that corresponds to 700
laser fringes of 1\,550\,nm). We have simulated the effect of this
loss on line broadening and estimated that it may add at most
10\,\kms{} to the global broadening.

We have estimated the amount of broadening added by the turbulent
motion and complex internal motion by comparing the expansion velocity
derived from our data (see section~\ref{sec:elliptical_prof}) to the
one measured by other authors and found that it lies between 10 and
15\,\kms{} for [\NII] and between 20 and 25\,\kms{} for H$\alpha$
(after subtraction of the thermal broadening and the instrumental
broadening). Note that turbulent motion cannot be separated from other
complex internal motions which would provide more than the simple two
emission lines we would expect for an ideal expanding shell. The
observation of more than two emission lines is reported in several
high-resolution studies \citepalias{Guerrero1997,
  Odell2013a}. Therefore, despite the small uncertainty of our
measurement, the derived value of the expansion velocity is an upper
limit that contains some unknown added quantity which can vary from
one pixel to another.

\begin{figure}
  \includegraphics[width=\linewidth]{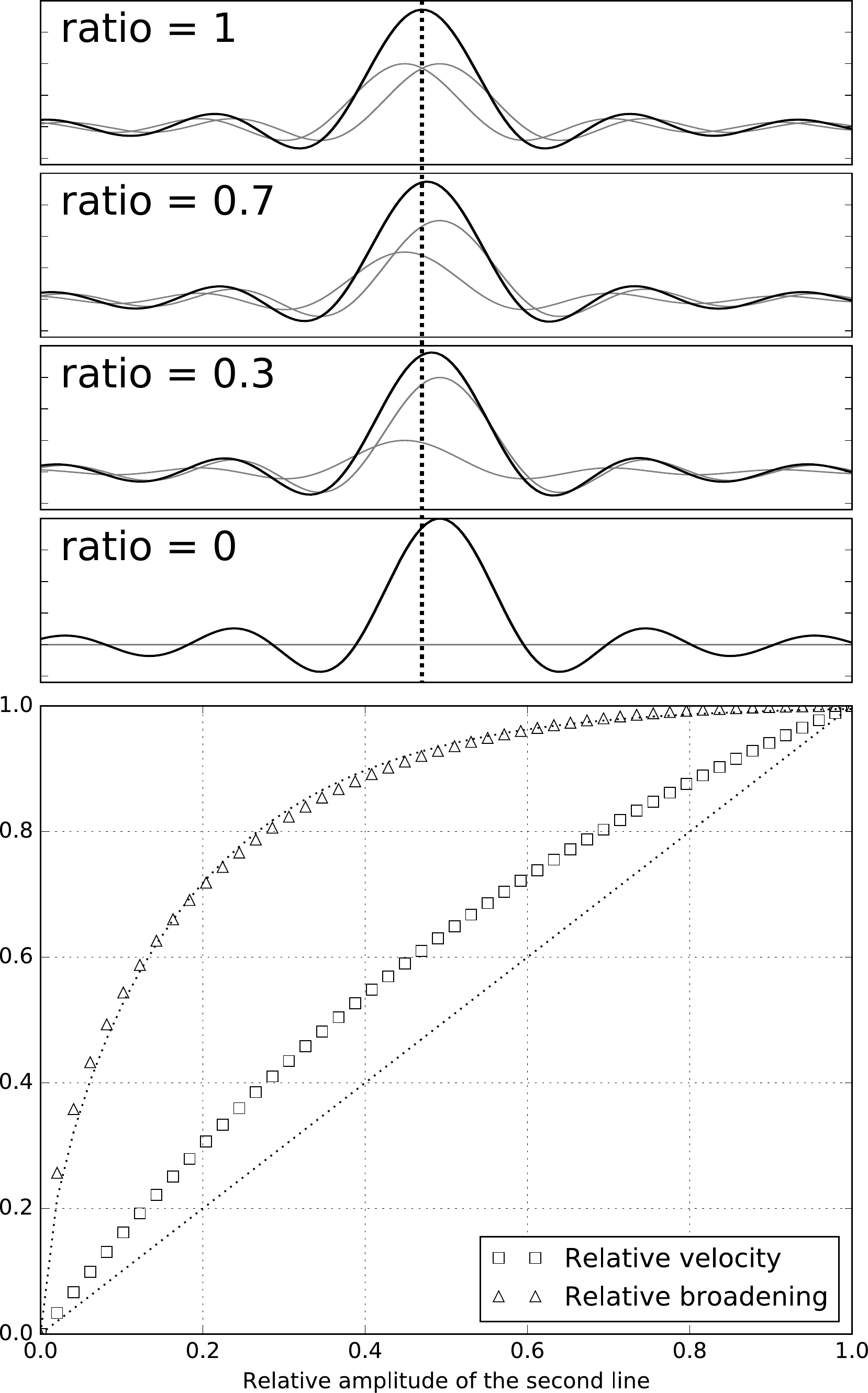}
  \caption{\textit{Top}: Sum of two unresolved H$\alpha$ lines
    observed at a resolution of 2600 created by an very thin shell
    expanding at a velocity of 40\,\kms{} for different intensity ratios
    of the lines. The real center of the two lines is indicated with
    a dotted line. \textit{Bottom}: Estimated velocity and broadening
    for different ratios of the intensity of two lines emitted by a
    thin shell expanding at a velocity of 40\,\kms{} observed at a
    resolution of 2600. An empirical fit of the broadening variation
    (see text for details) and the one-to-one line are indicated with
    dotted lines.}
  \label{fig:model2lines}
\end{figure}

We know that the line broadening is mainly caused by the
expansion of a thin shell, at least in the core and the Main Ring
regions. So that at a high resolution we should see two lines instead
of one separated by twice the expansion velocity. At low resolution
the resulting line can be treated as a broadened Gaussian line but
this model is only an approximation. First of all, we have checked
that this is a good approximation only if the line separation is less
than half the FWHM when the SNR is smaller than 1000 (e.g 60\,\kms{} at
a resolution of 2600). In this case the differences between the model
and the real spectrum are smaller than the noise (assuming a Gaussian
noise). Secondly, when the two emission lines do not have the same
intensity we can suspect that the centroid of the low-resolution line
will be shifted, causing a bias on the estimation of the line
velocity. There must also be some reduction of the measured
broadening. Fig.~\ref{fig:model2lines} shows the resulting
low-resolution line when the relative amplitude of the second line
changes. The expected line shift appears clearly. We can also have a
look at the bottom of Fig.~\ref{fig:model2lines} to see how the
measured velocity and broadening are affected.
We can notice that the velocity shift scales nearly linearly with the
amplitude ratio. But the broadening is much less affected. In fact it
starts to be underestimated by more than 10\,\% only when the amplitude
of the second line is 40\,\% the amplitude of the first line. An
empirical fit to this data gives an estimation of the relative error
$\Delta\sigma_{\text{Error}}$ for a given amplitude ratio $r$
\begin{equation}
  \Delta\sigma_{\text{Error}} \simeq 1 - \text{erf}(2\,r^{0.6})\;.
\end{equation}
We can therefore conclude that, at low resolution, the estimation of
the expansion velocity of a shell via the broadening of an emission
line will give reliable results while the measure of the velocity
itself, even at high SNR, is subject to a shift equal to half the
broadening. If we look at the core region of the [\NII] velocity map
in Fig.~\ref{fig:maps}, we can see a very large velocity difference
(around 30\,\kms{}) which is much smaller in the H$\alpha$ map. In fact
the front and the back of the nebula do not have the same intensity
along the line of sight and the centroid of the line is simply shifted
towards the most luminous portion of the shell. We are thus not
looking at a very disturbed velocity field at the center of
NGC\,6720. What the velocity map tells us here is probably that the
ionized material on the central shell is not homogeneously
distributed. Moreover the apparent high symmetry of the velocity field
suggests that the ejected material has been distributed with a strong
central symmetry with respect to the AGB star. The extracted velocity
maps must therefore be understood as a lower limit in the case of an
expanding shell where two lines are possibly present, but unresolved,
along the line of sight.

\subsubsection{3D velocity map of the Main Ring and the Central Lobes}

Given the limitations explained in the previous section, it appears
somehow naive to derive a 3d model of the [\NII] shell from our low
resolution data. But even if the resolution is lower than that of
echelle spectra, we must highlight the fact that it is the first time
velocity data is obtained with an homogeneous spatial sampling over
the entire nebula. It is interesting to compare what we can derive
from our data to the most complete and precise study of NGC\,6720 from
\cite{Odell2007} (henceforth \citetalias{Odell2007}) based on
long-slit spectroscopy at different position angles centered on
NGC\,6720. The high spectral resolution ($R \sim 42\,000$) of their
data is perfect to clearly separate the lines emitted from both sides
of the shell. But the spatial coverage is inhomogeneous and the data
must be interpolated between one position angle and the next, leading
to an increasing error as one moves away from the center of the slits.

\begin{figure}
  \includegraphics[width=\linewidth]{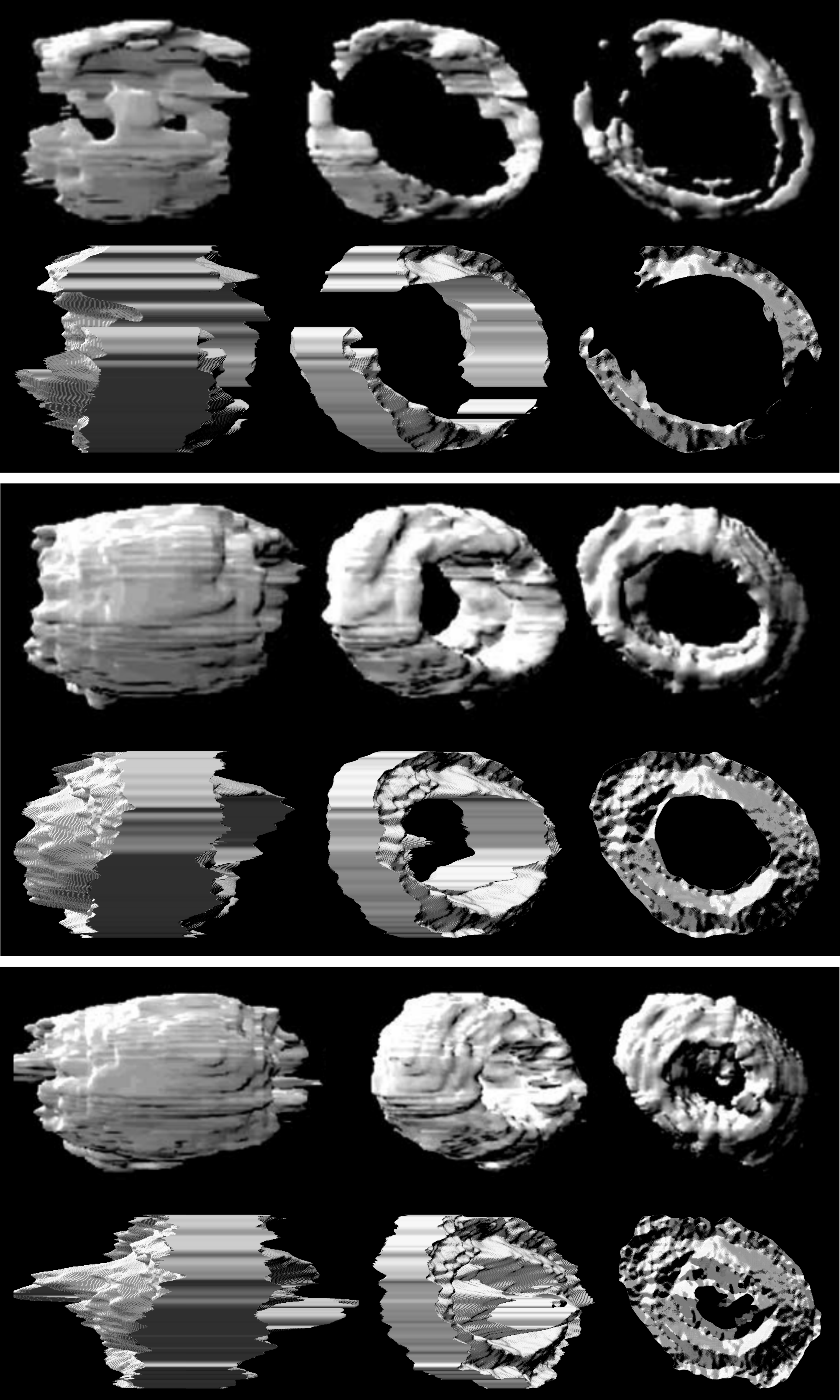}
  \caption{Opaque reconstruction determined from SITELLE's low
    resolution data (below) compared with \citetalias{Odell2007}
    (above) for 3 different flux levels (high, medium and low flux
    from top to bottom). Volumes are shown from three different
    directions: 90$^\circ$, 30$^\circ$ and along the line of sight
    from left to right. North is up and east to the left.}
  \label{fig:3d}
\end{figure}

We have constructed a 3D map of the velocity dispersion based on our
data at 3 different ranges of flux (high, medium and low surface
brightness) in order to mimic the figures shown in
\citetalias{Odell2007}. They are presented in Fig.~\ref{fig:3d}. Even
if the models are not probing the same parameters, they appear very
similar in many ways. The major difference can be seen at the boundary
of the region (especially on the lower intensity one) where the line
broadening, after a general decrease with an increasing radius, starts
to increase again at the boundary of the region. This general trend
also appears clearly in Fig.~\ref{fig:ellipsis} and is reported in
\citetalias{Guerrero1997}. The increasing broadening on the edges of
the main ring of the nebula gives a less spherical shape to our model
when compared to the model of \citetalias{Odell2007}. The region
affected corresponds to the inner halo of the nebula, where the
density decreases (see Fig.~\ref{fig:density} and Fig.~11 of
\citetalias{Odell2007}). The increasing broadening in this region
comes from a more complex velocity field due to the superposition of
the main ring shell and the halo shell. According to
\citetalias{Guerrero1997} the inner halo is the envelope from the
remnant red giant wind which has a continuous density distribution,
creating a broad component instead of two emission lines like the main
nebula shell. The analysis of \citetalias{Odell2013a} favors the
hypothesis of an expanding shell in the halo. This is also what
suggests our data (see Section~\ref{sec:halo}). If we look at the
Central Lobe, the general shape of the front surface is similar but it
appears exaggerated in our model. This certainly comes from the shift
of the centroid of the line that comes from the unequal surface
brightness of the two expanding fronts of the shell along the line of
sight. This effect is explained with more details in
Section~\ref{sec:limitations}.

\subsubsection{Elliptical profile}
\label{sec:elliptical_prof}

The inner region of the nebula encompasses a Central Lobe where the
emission is generally low and the [\NII]/H$\alpha$ ratio is smaller
than 0.5 and the Main Ring, the brightest part of the nebula where the
[\NII]/H$\alpha$ ratio is greater than 1 and goes up to 4 at its
boundary (see Fig.~\ref{fig:ratios} and \citetalias{Odell2013a}). 
\begin{figure}
  \includegraphics[width=\columnwidth]{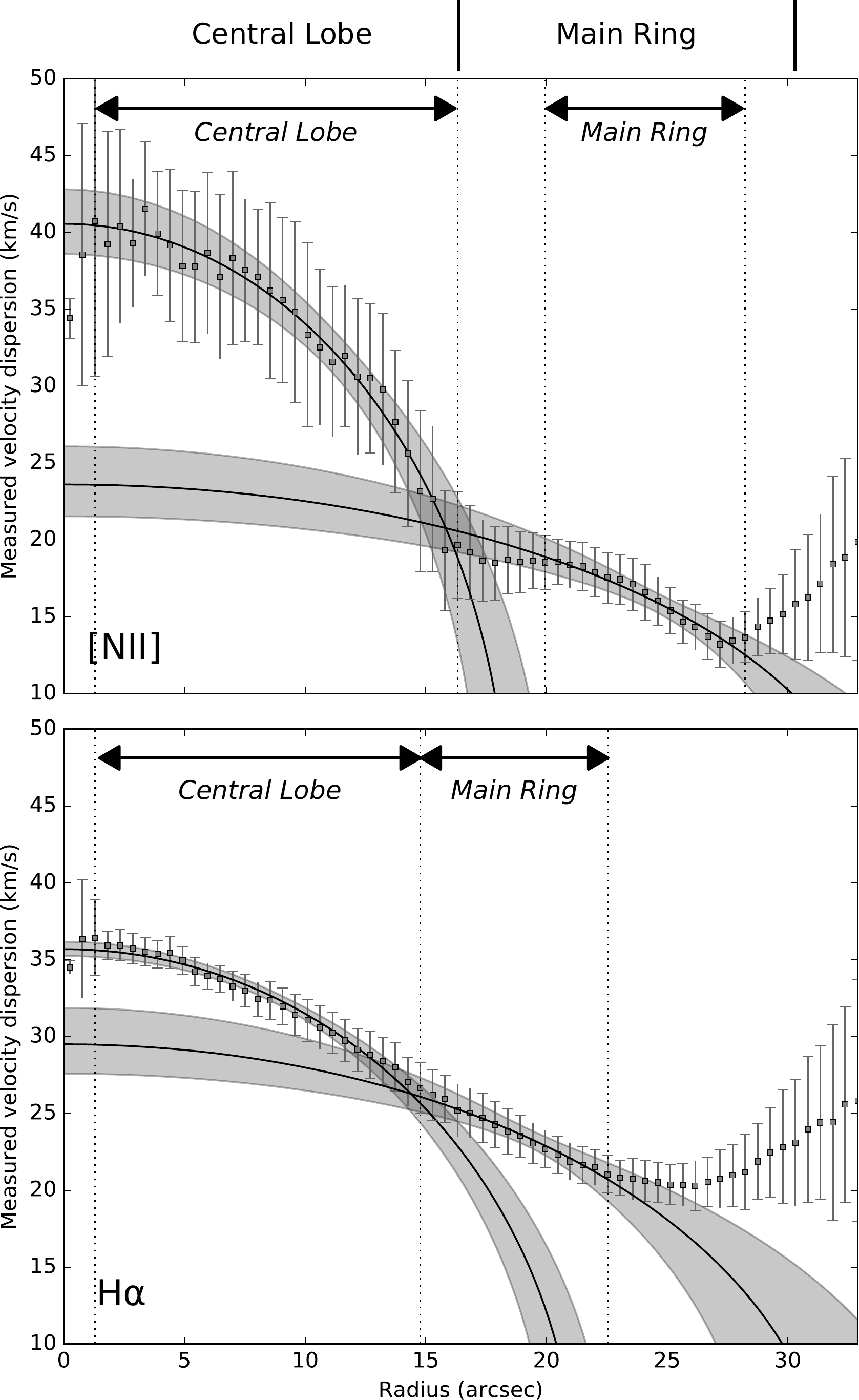}
  \caption{Elliptical profiles of the broadening of the [\NII] lines
    (\textit{top}) and the H$\alpha$ line (\textit{bottom}). The
    radius given is measured along the minor axis of the ellipses. The
    scale ellipse used to construct this figure corresponds to the
    Main Ring ellipse as shown in Fig.~\ref{fig:ratios}. For each
    profile, two elliptical expansion models have been fit (shown in
    solid black) with a Monte-Carlo Markov-Chain algorithm (see text
    for details). The error bars show the standard deviation of the
    broadening distribution along the ellipses. The regions used to
    fit the expansion models of the Central Lobe and the Main Ring are
    indicated with dotted lines. The regions defined by the models
    comprised between the 16th and 84th percentiles for each fit are
    shown in grey.}
  \label{fig:ellipsis}
\end{figure}
We have demonstrated in Section~\ref{sec:limitations} that the
broadening was not too much affected by the intensity ratio between
the two lines of an expanding shell. Therefore, its value gives a
reliable estimation of the radial velocity difference between these
lines. Given the fact that the nebula is seen nearly pole-on and
presents a good cylindrical symmetry along the polar axis, it is
possible to compute the elliptical profile of the broadening of [\NII]
and H$\alpha$. Those profiles are shown in
Fig.~\ref{fig:ellipsis}. The high effective temperature of the central
star creates a thin ionized front that is very well traced by the
[\NII] lines. If we look at the broadening profile of the [\NII] line,
the elliptical distribution of the radial velocity in the Central Lobe
appears clearly. The higher than expected standard deviation of its
value along each radius with respect to the small scatter of the
median values along the fitted elliptical profile is intriguing. This
is most certainly due for a good part to the effect of the large
difference of intensity of the two lines along the line of sight in
this region (see Section~\ref{sec:limitations}). Even if the
broadening value is less affected, we can see that it is not
negligible. The small scatter of the median value along the fitted
elliptical profile suggests that the expansion velocity of the ionized
region presents a very good central symmetry with respect to the
ionizing star, which therefore indicates a very good central symmetry
of the physical properties (temperature, density and abundances) of
the medium crossed by the ionization front since its origins. The same
central symmetry must have then be respected by the oldest AGB winds
which remnants constitute the ionized material of the Outer Halo (see
Section~\ref{sec:halo}). Also, the presence of two shells expanding at
different velocities, already deduced by \citetalias{Odell2013a}, can
be confirmed. The faster shell corresponds to the central core of the
nebula where the density is smaller while the slower one corresponds
to the main ring where the density is much higher (see
Fig.~\ref{fig:ratios}).

An elliptical model has been fit to the data to compute the expansion
velocity of each shell. If we consider the expansion velocity to be
the same in all directions then a pure ellipsis is a perfect
model. But we have to add to this model a constant velocity component
which comes from thermal broadening, turbulence and the dispersion of
the expansion velocity around its mean value.
\begin{equation}
  \sigma = \sqrt{\sigma_{\text{T}}^2 + \sigma_{\text{V}}^2+ V_{\text{Exp}}^2\left(1 - \frac{r}{r_0}\right)^2}\;,
\end{equation}
where $\sigma$ is the measured broadening, $V_{\text{Exp}}$, the
expansion velocity, $\sigma_{\text{T}}$, the thermal broadening,
$\sigma_{\text{V}}$, the dispersion of the velocity coming from other
origins, $r$ is the radius from the center and $r_0$ the radial scale
parameter. We have tried to fit such a model with a Markov Chain
Monte-Carlo method but the lack of data at the very edges of the
shell, where the relative importance of the other sources of
broadening becomes significant, does not permit to fit a model with so
many parameters. Occam's razor clearly gives the preeminence to the
simplest model:
\begin{equation}
  \sigma = \sqrt{V_{\text{Exp}}^2\left(1 - \frac{r}{r_0}\right)^2 + \sigma_{\text{T}}^2}\;,
\end{equation}

With this method, the fitted expansion velocity is 40.5\,$\pm
2.2$\,\kms{} for the central lobe and 23.5\,$\pm2.2$\,\kms{} for the
main ring which are both a few \kms{} above the expansion velocities
measured by \citetalias{Odell2013a} (the given velocities are
corrected for the estimation of the thermal broadening of [\NII] at
10\,000\,K). Note that only the outer part of the main ring has been
used for the fit as the inner portion does not follow an elliptical
shape. The expansion velocity of the PDR follows very well the density
profile (see Fig.~\ref{fig:ratios}) suggesting that the ionization
front is moving faster in low density regions.

If we look at the fit made for the H$\alpha$ broadening, a simple
elliptical model fails to explain its really homogeneous disk-like
shape. This most certainly reflects the fact that hydrogen is ionized
over a much deeper layer than [\NII], thus smoothing the velocity
differences between the Main Ring and the Central Lobe. The sum of two
elliptical shells better explains the overall profile of the H$\alpha$
broadening. The fitted expansion velocity in the core, corrected for
the estimation of the thermal broadening of H$\alpha$ at 10\,000\,K,
is 34.5\,$\pm0.5$\,\kms{}. The fitted expansion velocity of the
outer shell that corresponds roughly to the Main Ring is
28.1\,$\pm2.2$\,\kms{}.

\subsection{Study of the Halo}
\label{sec:halo}

As \citetalias{Odell2013a} suggested, ``the radial velocity of the
Halo features are difficult to determine because of their low surface
brightness''. Up to now, the only data available with enough
signal-to-noise ratio comes from the long slit study of
\citetalias{Guerrero1997}. Their Fig.~8 shows a broad component that
narrows from the Inner Halo to the edge of the Outer
Halo. \citetalias{Odell2013a} see in this figure a velocity ellipse
corresponding to an expansion velocity of 15\,\kms{}. Despite the low
resolution of our spectra, its high spatial resolution combined with
an integration time of 2.5 hours give us the deepest and most
complete picture of the Halo up to date. A general examination of the
[\NII] velocity map in Fig.~\ref{fig:maps} shows that the Inner Halo
is composed of two ionized fronts with a velocity difference of more
than 30\,\kms{}. These two fronts are obvious when we look out of the
Main Ring but near the Main Ring, the dim petals of the Inner Halo are
likely to be hidden by its very high surface brightness. The two
fronts are similar in shape but they appear shifted with respect to
one another along the minor axis of the nebula: one front is expanding
towards us in the SW region while the other is expanding away from us
in the NE direction. According to \citetalias{Odell2013a}, the nebula is
supposed to be tilted by 5$^{\circ}$ along the minor axis which
creates a small velocity difference of 5\,\kms{}. The velocity
difference between the two fronts thus cannot be the result of this
tilt. Moreover, the front expanding away from us is also visible in the
SW region. This suggests that the Inner Halo is a shell expanding in
an inhomogeneous medium with a strong bipolar symmetry. The two poles
of the ionized front are at least roughly aligned with the polar axis
of central region of the nebula. The expansion velocity must be at
least of 15\,\kms{} but this is a lower limit because the expansion
velocity cannot be measured along the line of sight. The relative
velocity of the whole shell with respect to the Local Standard of Rest
is around -15\,\kms{}. In the regions where the two fronts are
overlapping the measured velocity is also around this value. This
effect which comes from the mix of two emission lines with a similar
intensity is explained in details in
Section~\ref{sec:limitations}. The arcuate emission features are
coming from the complex morphology of the ionized front composed of
portions of bubbles.

The small ionized regions lying on the edge of the Outer Halo are too
dim to be studied with a good spatial resolution. We have thus used
integrated spectra of small circular regions to obtain a more precise
measurement of their velocity (see
Fig.~\ref{fig:maps}). Interestingly, most of the Outer Halo features
have an LSR velocity around -20\,\kms{}, that is very near the mean
velocity of the Inner Halo bipolar shell. This suggests that the Inner
Halo and the Outer Halo may be in fact the same shell. This shell
would be made of a two bright poles. The dim and inhomogeneous
emission regions distributed along a circle can be seen because
of a projection effect at the boundaries of the Halo shell.

The overall symmetry of the ejection of material in the core and the
Halo is remarkable. The velocity map suggests that the SW part of the
core is filled with material moving away from us while the SW part of
the Halo is filled with material moving away from us. The exact
counterpart of these opposite ejections in terms of volume and shape
can be seen in the NE part of the core and the Halo.

\section{Conclusions}

We have defined a better implementation of a sinc instrumental line
convolved with a Gaussian profile and have designed the corresponding
set of Gaussian apodizing functions that are now used by the reduction
pipeline ORBS. We have used this model for the analysis of the
commissioning data obtained with SITELLE at CFHT that covers the
planetary nebula NGC\,6720. Our major conclusions are listed below.
\begin{enumerate}
\item Our implementation of the convolution of a sinc and a Gaussian
  profile is much better at fitting small broadening values.
\item A complete spectrum model consisting of multiple lines and a
  low-order continuum can be fitted to the spectra of NGC\,6720 and
  respects the underlying hypotheses of the Levenberg-Marquardt
  minimization algorithm.
\item We have demonstrated that the broadening of an unresolved line
  emitted by a shell could give a robust estimation of its expansion
  velocity while the position of its centroid was greatly affected by
  the intensity ratio of the two lines emitted by both fronts of the
  shell along the line of sight.
\item We have derived the first complete density map of the central
  part of the nebula which clearly shows that the Main Ring is the
  region of highest density.
\item The study of the broadening of the [\NII] lines have
  clearly demonstrated, as \citetalias{Odell2013a} originally
  suggested, that the Main Ring and the Central Lobe are two different
  shells with different expansion velocities.
\item We have derived the deepest and the most spatially resolved
  velocity maps of the Halo in [\NII] and H$\alpha$.
\item The study of the velocity of the Halo features reveals that the
  brightest bubbles are originating from two bipolar structures
  with a velocity difference of more than 35\,\kms{} lying at the poles
  of a possibly unique Halo shell expanding at a velocity of more than
  15\,\kms{}.
\end{enumerate}
The great quality and the quantity of data obtained with SITELLE on
this object reveals much more on this nebula than the few aspects
covered here. The investigation of these aspects will be published in
a next article. The observation with SITELLE of the lines
[\OIII]$\lambda\lambda$4959,5007, [\OIII]$\lambda$4363, the
[\OII]$\lambda$3727 doublet and the [\NII]$\lambda$5755 line will
certainly permit a much deeper and more precise study of this object.


\section*{Acknowledgements}
Based on observations obtained with SITELLE, a joint project of
Universit{\'e} Laval, ABB, Universit{\'e} de Montr{\'e}al and the
Canada-France-Hawaii Telescope (CFHT) which is operated by the
National Research Council (NRC) of Canada, the Institut National des
Science de l'Univers of the Centre National de la Recherche
Scientifique (CNRS) of France, and the University of Hawaii. LD is
grateful to the Natural Sciences and Engineering Research Council of
Canada, the Fonds de Recherche du Qu{\'e}bec, and the Canadian
Foundation for Innovation for funding.





\bibliographystyle{mnras}
\bibliography{sincfit} 

\begin{thebibliography}{}
\makeatletter
\relax
\def\mn@urlcharsother{\let\do\@makeother \do\$\do\&\do\#\do\^\do\_\do\%\do\~}
\def\mn@doi{\begingroup\mn@urlcharsother \@ifnextchar [ {\mn@doi@}
  {\mn@doi@[]}}
\def\mn@doi@[#1]#2{\def\@tempa{#1}\ifx\@tempa\@empty \href
  {http://dx.doi.org/#2} {doi:#2}\else \href {http://dx.doi.org/#2} {#1}\fi
  \endgroup}
\def\mn@eprint#1#2{\mn@eprint@#1:#2::\@nil}
\def\mn@eprint@arXiv#1{\href {http://arxiv.org/abs/#1} {{\tt arXiv:#1}}}
\def\mn@eprint@dblp#1{\href {http://dblp.uni-trier.de/rec/bibtex/#1.xml}
  {dblp:#1}}
\def\mn@eprint@#1:#2:#3:#4\@nil{\def\@tempa {#1}\def\@tempb {#2}\def\@tempc
  {#3}\ifx \@tempc \@empty \let \@tempc \@tempb \let \@tempb \@tempa \fi \ifx
  \@tempb \@empty \def\@tempb {arXiv}\fi \@ifundefined
  {mn@eprint@\@tempb}{\@tempb:\@tempc}{\expandafter \expandafter \csname
  mn@eprint@\@tempb\endcsname \expandafter{\@tempc}}}

\bibitem[\protect\citeauthoryear{Baril et~al.,}{Baril et~al.}{2016}]{Baril2016}
Baril M.~R.,  et~al., 2016, in Evans C.~J.,  Simard L.,   Takami H.,  eds,
  Proceedings of SPIE. International Society for Optics and Photonics, p.
  990829, \mn@doi{10.1117/12.2232075}

\bibitem[\protect\citeauthoryear{Bernier}{Bernier}{2006}]{Bernier2006}
Bernier A.-P.,  2006, in Proceedings of SPIE. SPIE, pp 626949--626949--9,
  \mn@doi{10.1117/12.671410}

\bibitem[\protect\citeauthoryear{Bohlin, Dickinson  \& Calzetti}{Bohlin
  et~al.}{2001}]{Bohlin2001}
Bohlin R.~C.,  Dickinson M.~E.,   Calzetti D.,  2001, \mn@doi [The Astronomical
  Journal] {10.1086/323137}, 122, 2118

\bibitem[\protect\citeauthoryear{Davis, Abrams  \& Brault}{Davis
  et~al.}{2001}]{Davis2001}
Davis S.~P.,  Abrams M.~C.,   Brault J. W. J.~W.,  2001, {Fourier transform
  spectrometry}.
Academic Press, San Diego

\bibitem[\protect\citeauthoryear{Drissen, Bernier, Rousseau-Nepton, Alarie,
  Robert, Joncas, Thibault  \& Grandmont}{Drissen et~al.}{2010}]{Drissen2010}
Drissen L.,  Bernier A.-P.,  Rousseau-Nepton L.,  Alarie A.,  Robert C.,
  Joncas G.,  Thibault S.,   Grandmont F.,  2010, in McLean I.~S.,  Ramsay
  S.~K.,   Takami H.,  eds, ~ Vol. 7735, SPIE Astronomical Telescopes +
  Instrumentation. International Society for Optics and Photonics, pp
  77350B--77350B--10, \mn@doi{10.1117/12.856470}

\bibitem[\protect\citeauthoryear{Foreman-Mackey}{Foreman-Mackey}{2016}]{Foreman2016}
Foreman-Mackey D.,  2016, \mn@doi [The Journal of Open Source Software]
  {10.21105/joss.00024}, 24

\bibitem[\protect\citeauthoryear{Foreman-Mackey, Hogg, Lang  \&
  Goodman}{Foreman-Mackey et~al.}{2012}]{Foreman-Mackey2012}
Foreman-Mackey D.,  Hogg D.~W.,  Lang D.,   Goodman J.,  2012, \mn@doi
  [Publications of the Astronomical Society of Pacific, Volume 125, Issue 925,
  pp. 306-312 (2013).] {10.1086/670067}, 125, 306

\bibitem[\protect\citeauthoryear{Formisano et~al.,}{Formisano
  et~al.}{2005}]{Formisano2005}
Formisano V.,  et~al., 2005, \mn@doi [Planetary and Space Science]
  {10.1016/j.pss.2004.12.006}, 53, 963

\bibitem[\protect\citeauthoryear{Grandmont}{Grandmont}{2003}]{Grandmont2003}
Grandmont F.,  2003, in Proceedings of SPIE. SPIE, pp 392--401,
  \mn@doi{10.1117/12.457339}

\bibitem[\protect\citeauthoryear{Griffin et~al.,}{Griffin
  et~al.}{2010}]{Griffin2010}
Griffin M.~J.,  et~al., 2010, \mn@doi [Astronomy and Astrophysics]
  {10.1051/0004-6361/201014519}, 518, L3

\bibitem[\protect\citeauthoryear{Guerrero, Manchado, Chu, Uerrero, Anchado  \&
  Hu}{Guerrero et~al.}{1997}]{Guerrero1997}
Guerrero M. A. M.~A.,  Manchado A.,  Chu Y.-H.,  Uerrero M. A.~G.,  Anchado
  A.~M.,   Hu Y.~C.,  1997, \mn@doi [The Astrophysical Journal]
  {10.1086/304582}, 487, 328

\bibitem[\protect\citeauthoryear{Harris}{Harris}{1978}]{Harris1978}
Harris F.,  1978, \mn@doi [Proceedings of the IEEE] {10.1109/PROC.1978.10837},
  66, 51

\bibitem[\protect\citeauthoryear{Jones, Oliphant, Peterson  et~al.}{Jones
  et~al.}{2001}]{Jones2001}
Jones E.,  Oliphant T.,  Peterson P.,   et~al., 2001, {SciPy}: Open source
  scientific tools for {Python}, \url {http://www.scipy.org/}

\bibitem[\protect\citeauthoryear{Kauppinen \& Partanen}{Kauppinen \&
  Partanen}{2001}]{Kauppinen2001}
Kauppinen J.,  Partanen J.,  2001, {Fourier Transforms in Spectroscopy}.
Wiley-VCH

\bibitem[\protect\citeauthoryear{Kawada et~al.,}{Kawada
  et~al.}{2008}]{Kawada2008}
Kawada M.,  et~al., 2008, Publications of the Astronomical Society of Japan

\bibitem[\protect\citeauthoryear{Kunde et~al.,}{Kunde et~al.}{1996}]{Kunde1996}
Kunde V.~G.,  et~al., 1996, Proc. SPIE Vol. 2803, 2803, 162

\bibitem[\protect\citeauthoryear{Levenberg}{Levenberg}{1944}]{Levenberg1944}
Levenberg K.,  1944, Quarterly of Applied Mathematics, 2, 164

\bibitem[\protect\citeauthoryear{Luridiana, Morisset  \& Shaw}{Luridiana
  et~al.}{2012}]{Luridiana2012}
Luridiana V.,  Morisset C.,   Shaw R.~A.,  2012, in Planetary Nebulae: An Eye
  to the Future. IAU Symposium, pp 422--423, \mn@doi{10.1017/S1743921312011738}

\bibitem[\protect\citeauthoryear{Maillard \& Michel}{Maillard \&
  Michel}{1982}]{Maillard1982}
Maillard J.,  Michel G.,  1982, {Instrumentation for Astronomy with Large
  Optical Telescopes}.
 Astrophysics and Space Science Library Vol. 92, Springer Netherlands,
  Dordrecht, \mn@doi{10.1007/978-94-009-7787-7}

\bibitem[\protect\citeauthoryear{Marquardt}{Marquardt}{1963}]{Marquardt1963}
Marquardt D.~W.,  1963, \mn@doi [Journal of the Society for Industrial and
  Applied Mathematics] {10.1137/0111030}, 11, 431

\bibitem[\protect\citeauthoryear{Martin}{Martin}{2015}]{Martin2015-thesis}
Martin T.,  2015, Phd thesis, Universit\'{e} Laval

\bibitem[\protect\citeauthoryear{Martin, Drissen  \& Joncas}{Martin
  et~al.}{2012}]{Martin2012}
Martin T.,  Drissen L.,   Joncas G.,  2012, in Radziwill N.~M.,  Chiozzi G.,
  eds, ~ Vol. 2, SPIE - Software and Cyberinfrastructure for Astronomy II. pp
  84513K--84513K--9, \mn@doi{10.1117/12.925420}

\bibitem[\protect\citeauthoryear{Martin, Drissen  \& Joncas}{Martin
  et~al.}{2015}]{Martin2015}
Martin T.,  Drissen L.,   Joncas G.,  2015, Astronomical Data Analysis Software
  an Systems XXIV (ADASS XXIV), 495

\bibitem[\protect\citeauthoryear{Medina, Arthur, Henney, Mellema  \&
  Gazol}{Medina et~al.}{2014}]{medina2014}
Medina S.-N.~X.,  Arthur S.~J.,  Henney W.~J.,  Mellema G.,   Gazol a.,  2014,
  \mn@doi [Monthly Notices of the Royal Astronomical Society]
  {10.1093/mnras/stu1862}, 445, 1797

\bibitem[\protect\citeauthoryear{Naylor \& Tahic}{Naylor \&
  Tahic}{2007}]{Naylor2007}
Naylor D.~A.,  Tahic M.~K.,  2007, \mn@doi [Journal of the Optical Society of
  America A] {10.1364/JOSAA.24.003644}, 24, 3644

\bibitem[\protect\citeauthoryear{Naylor, Gom  \& Zhang}{Naylor
  et~al.}{2006}]{Naylor2006}
Naylor D.~A.,  Gom B.~G.,   Zhang B.,  2006, in Zmuidzinas J.,  Holland W.~S.,
  Withington S.,   Duncan W.~D.,  eds, ~ Vol. 6275, Millimeter and
  Submillimeter Detectors and Instrumentation for Astronomy III. Edited by
  Zmuidzinas. pp 62751Z--62751Z--11, \mn@doi{10.1117/12.670552}

\bibitem[\protect\citeauthoryear{O'Dell, Sabbadin  \& Henney}{O'Dell
  et~al.}{2007}]{Odell2007}
O'Dell C.~R.,  Sabbadin F.,   Henney W.~J.,  2007, \mn@doi [The Astronomical
  Journal] {10.1086/521823}, 134, 1679

\bibitem[\protect\citeauthoryear{O'Dell, Ferland, Henney  \& Peimbert}{O'Dell
  et~al.}{2013a}]{Odell2013a}
O'Dell C.~R.,  Ferland G.~J.,  Henney W.~J.,   Peimbert M.,  2013a, \mn@doi
  [The Astronomical Journal] {10.1088/0004-6256/145/4/92}, 145, 92

\bibitem[\protect\citeauthoryear{O'Dell, Ferland, Henney  \& Peimbert}{O'Dell
  et~al.}{2013b}]{Odell2013c}
O'Dell C.~R.,  Ferland G.~J.,  Henney W.~J.,   Peimbert M.,  2013b, \mn@doi
  [The Astronomical Journal] {10.1088/0004-6256/145/4/93}, 145, 93

\bibitem[\protect\citeauthoryear{O'dell}{O'dell}{1991}]{Odell1991}
O'dell C.~R.,  1991, in Fragmentation of Molecular Clouds and Star Formation.
  Springer Netherlands, Dordrecht, pp 476--479,
  \mn@doi{10.1007/978-94-011-3384-5_76}

\bibitem[\protect\citeauthoryear{Osterbrock \& Ferland}{Osterbrock \&
  Ferland}{2006}]{OsterbrockDonaldE.2006}
Osterbrock D.,  Ferland G.,  2006, {Astrophysics of gaseous nebulae and active
  galactic nuclei}.
University Science Books, Sausalito

\bibitem[\protect\citeauthoryear{Ott}{Ott}{2010}]{Ott2010}
Ott S.,  2010, Astronomical Data Analysis Software and Systems XIX (ADASS XIX),
  434

\bibitem[\protect\citeauthoryear{Shaw \& Dufour}{Shaw \&
  Dufour}{1995}]{Shaw1995}
Shaw R.~A.,  Dufour R.~J.,  1995, \mn@doi [Publications of the Astronomical
  Society of the Pacific] {10.1086/133637}, 107, 896

\bibitem[\protect\citeauthoryear{Wright, Corradi  \& Perinotto}{Wright
  et~al.}{2005}]{Wright2005}
Wright S.~a.,  Corradi R. L.~M.,   Perinotto M.,  2005, \mn@doi [Astronomy and
  Astrophysics] {10.1051/0004-6361:20052666}, 436, 9

\makeatother
\end{thebibliography}





\appendix

\section{Demonstration of the Gaussian-sinc convolution formula}
\label{appendix_demo}

Even if the convolution of a sinc and a Gaussian is a result that can
easily be found in the literature, the demonstration of the function
considered in this paper (see equation~\ref{eq:erf_form}), which is
not trivial, is not given anywhere. Since only some of the major steps
can be found in the Herschel Common Science System (HCSS)
documentation, we provide here the full demonstration for reference.

Starting from equation~\ref{eq:convolve} and following the convolution
theorem, the convolution of a sinc and a Gaussian $SG(\sigma)$ can be
rewritten as the inverse Fourier transform of
\subsection{Analytic form of the convolution of a sinc and a Gaussian}

the product of the
Fourier transforms of the convoluted function which leads to
\begin{equation}
  SG(\sigma)=\int_{-\infty}^{+\infty} \hat{S}(x) \hat{\delta}(x) \hat{G}(x)\,e^{2i\pi \sigma x}\,\text{d}x\;.
\end{equation}
Let's first write down the Fourier transform of the 3 components of
$SG(\sigma)$ as they can be found, for example, in
\cite{Kauppinen2001}:
\begin{equation}
  \hat{S}(x)=\pi \Delta w \;\Pi\left(x\pi \Delta w\right) = 
  \begin{cases} 
    \pi\Delta w & \text{if } |x| \leq \frac{1}{2\pi\Delta w} \\
    0       & \text{otherwise}
  \end{cases}\;,
\end{equation}
 
\begin{gather}
  \hat{\delta}(x)=e^{-2i\pi \sigma_0 x}\,,\\
  \hat{G}(x)=\sqrt{2\pi} \Delta\sigma e^{-2\Delta\sigma^2\pi^2 x^2}\;.
\end{gather}
It follows that
\begin{multline}
  SG(\sigma)= \sqrt{2\pi}\Delta\sigma\pi\Delta w\\\int_{-\infty}^{+\infty} \Pi\left(x\pi \Delta w\right) e^{-2\Delta\sigma^2\pi^2 x^2} e^{2i\pi x (\sigma - \sigma_0)} \,\text{d}x\;.
\end{multline}
The rectangular function $\Pi$ defines new boundaries to the integral
and the Euler's formula permit to separate this integral into an even
term (multiplied by a $\cos$) and an odd term (multiplied by a
$\sin$)
\begin{multline}
  SG(\sigma)=\sqrt{2\pi}\Delta\sigma\pi\Delta w  \bigg[\int_{-\frac{1}{2\pi \Delta w}}^{+\frac{1}{2\pi \Delta w}}  e^{-2\Delta\sigma^2\pi^2 x^2}  \cos({2\pi x(\sigma - \sigma_0)}) \,\text{d}x \\
  + i \int_{-\frac{1}{2\pi \Delta w}}^{+\frac{1}{2\pi \Delta w}}  e^{-2\Delta\sigma^2\pi^2 x^2} \sin({2\pi x(\sigma - \sigma_0)}) \,\text{d}x\bigg]\;.
\end{multline}
The integral of the odd term is equal to 0 and the integral of the even term is simply twice the integral along the positive part. Using once again Euler's formula we can split the even term into two exponential terms
\begin{multline}
  SG(\sigma)=\sqrt{2\pi}\Delta\sigma\pi\Delta w  \bigg[\int_{0}^{\frac{1}{2\pi \Delta w}}  e^{2i\pi x (\sigma - \sigma_0) - 2\Delta\sigma^2\pi^2 x^2} \,\text{d}x \\
  + \int_{0}^{+\frac{1}{2\pi \Delta w}} e^{- 2i\pi x (\sigma - \sigma_0) - 2\Delta\sigma^2\pi^2 x^2} \,\text{d}x\bigg]\;.
\end{multline}
We can rearrange this equation by completing the square into both terms and rewrite it as
\begin{multline}
  SG(\sigma)=\sqrt{2\pi}\Delta\sigma\pi\Delta w  \exp\left({-\frac{(\sigma - \sigma_0)^2}{2\Delta\sigma^2}}\right) \bigg[\int_{0}^{z}  e^{-\alpha (x + i\beta)^2} \,\text{d}x \\
  + \int_{0}^{z} e^{-\alpha (x + i\beta)^2} \,\text{d}x\bigg]\;.
\end{multline}
where $\alpha = 2\Delta\sigma^2\pi^2$, $\beta = \cfrac{\sigma -
  \sigma_0}{2\pi\Delta\sigma^2}$ and $z = \cfrac{1}{2\pi \Delta
  w}$. 

The first integral can be found by substituting $x$ with $u =
x+i\beta$ 
\begin{alignat}{1}
  \int_{0}^{z}  e^{-\alpha (x + i\beta)^2} \,\text{d}x = \int_{ib}^{z+ib}  e^{-\alpha u^2} \,\text{d}u\\ = \int_{0}^{z+ib}  e^{-\alpha u^2} \,\text{d}u - \int_{0}^{ib}  e^{-\alpha u^2} \,\text{d}u\;.
\end{alignat}
From the erf definition,
\begin{equation}
  \int_0^z e^{-ax^2}=\frac{\sqrt{\pi}}{2\sqrt{a}}\,\text{erf}\left(\sqrt{a}z\right)\;,
\end{equation}
the value of the first integral becomes,
\begin{alignat}{1}
  \int_{0}^{z}  e^{-\alpha (x + i\beta)^2} \,\text{d}x = \frac{\sqrt{\pi}}{2\sqrt{\alpha}}\,\left(\text{erf}\left(\sqrt{\alpha}(z+i\beta)\right) - \text{erf}\left(i\sqrt{\alpha}\beta\right)\right)\,.
\end{alignat}
In an analogous fashion we can substitute $x$ with $u = x-i\beta$ in the second integral and rewrite it as
\begin{alignat}{1}
  \int_{0}^{z}  e^{-\alpha (x - i\beta)^2} \,\text{d}x = \frac{\sqrt{\pi}}{2\sqrt{\alpha}}\,\left(\text{erf}\left(\sqrt{\alpha}(z-i\beta)\right) + \text{erf}\left(i\sqrt{\alpha}\beta\right)\right)\;.
\end{alignat}
$SG(\sigma)$ is therefore
\begin{multline}
  SG(\sigma)=\frac{\pi\Delta w}{2} \exp\left({-\frac{(\sigma - \sigma_0)^2}{2\Delta\sigma^2}}\right) \bigg[\text{erf}\left(\frac{\Delta\sigma}{\sqrt{2}\Delta w} + i\frac{\sigma - \sigma_0}{\sqrt{2}\Delta\sigma}\right)\\
+ \text{erf}\left(\frac{\Delta\sigma}{\sqrt{2}\Delta w} - i\frac{\sigma - \sigma_0}{\sqrt{2}\Delta\sigma}\right)\bigg]\;.
\label{eq:unnorm_sg}
\end{multline}

\subsection{Normalized formula}

Now we might want to normalize the above function so that its peak value (when $\sigma = \sigma_0$) is 1, i.e. $SG(\sigma_0)=1$. This translates to
\begin{equation}
  \pi\Delta w\; \text{erf}\left(\frac{\Delta\sigma}{\sqrt{2}\Delta w}\right) =1\;,
\end{equation}
which permits to write the normalized convolution of a sinc and a
Gaussian given in equation~\ref{eq:erf_form} as
\begin{multline}
  SG(\sigma)=\frac{\exp\left({-\frac{(\sigma - \sigma_0)^2}{2\Delta\sigma^2}}\right)}{2 \text{erf}\left(\frac{\Delta\sigma}{\sqrt{2}\Delta w}\right)} \bigg[\text{erf}\left(\frac{\Delta\sigma}{\sqrt{2}\Delta w} + i\frac{\sigma - \sigma_0}{\sqrt{2}\Delta\sigma}\right)\\
+ \text{erf}\left(\frac{\Delta\sigma}{\sqrt{2}\Delta w} - i\frac{\sigma - \sigma_0}{\sqrt{2}\Delta\sigma}\right)\bigg]\;.
\label{eq:norm_sg}
\end{multline}

\subsection{Value of the integral on the whole space}

As $S(\sigma)$, $\delta(\sigma - \sigma_0)$ and $G(\sigma)$ are
integrable functions, the integral of $SG(\sigma)$ on the whole space
can be computed easily if we consider, after the Fubini's theorem,
that it is the product of the integrals of $S(\sigma)$, $\delta(\sigma
- \sigma_0)$ and $G(\sigma)$. In the case of its unnormalized version
(see equation~\ref{eq:unnorm_sg}) it follows
\begin{gather}
  \int_{-\infty}^{+\infty} SG(\sigma) = \int_{-\infty}^{+\infty} S(\sigma) * \delta(\sigma - \sigma_0) * G(\sigma) 
\\= \int_{-\infty}^{+\infty} S(\sigma) \times \int_{-\infty}^{+\infty} \delta(\sigma - \sigma_0) \times \int_{-\infty}^{+\infty} G(\sigma)
\\= \sqrt{2\pi}\Delta\sigma \times \pi\Delta w\,.
\end{gather}
The integral of the normalized function (see
equation~\ref{eq:norm_sg}) can be obtained by dividing this results
by the normalization factor, therefore
\begin{gather}
  \int_{-\infty}^{+\infty} SG(\sigma) = \frac{\sqrt{2\pi}\Delta\sigma}{\text{erf}\left(\frac{\Delta\sigma}{\sqrt{2}\Delta w}\right)}\,.
\end{gather}

\section{Absolute velocity calibration of NGC\,6720}
\label{sec:abs_vel_calib}

\begin{figure*}
  \includegraphics[width=\linewidth]{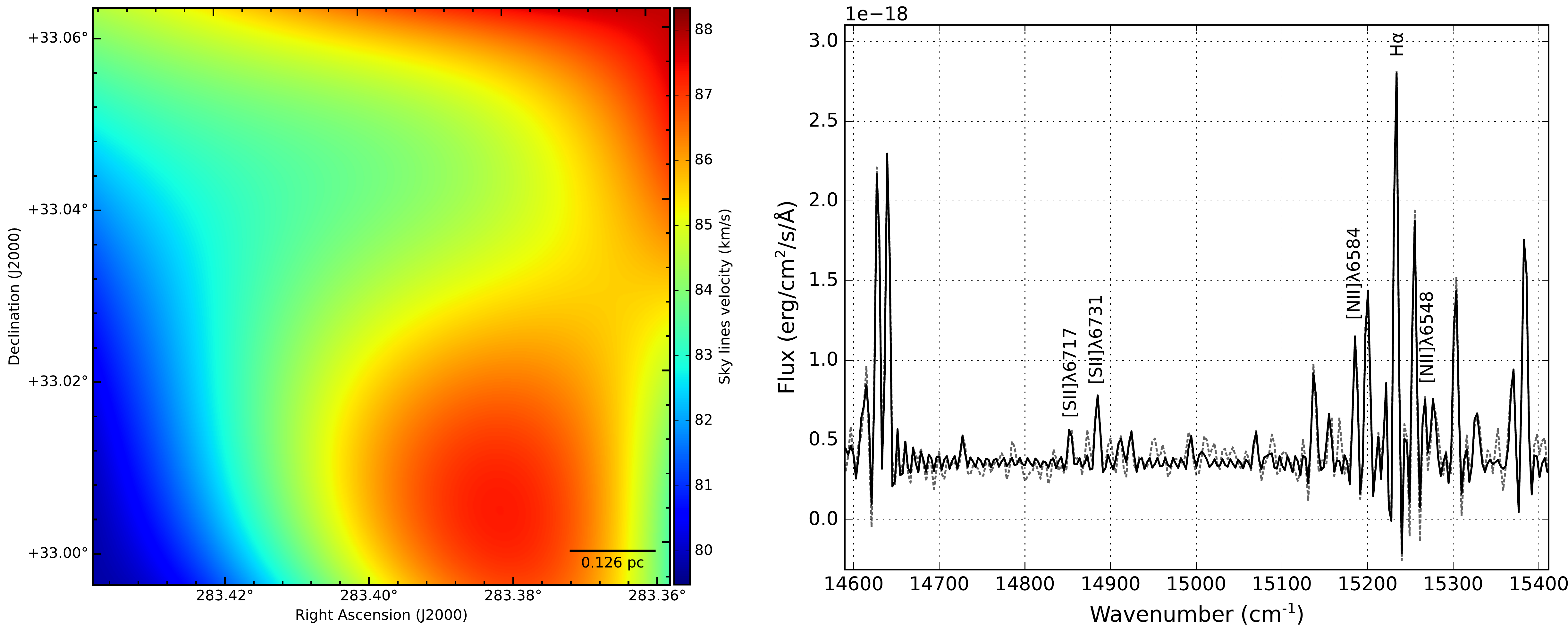}
  \caption{\textit{Left}: Map of the measured sky lines velocity used
    to calibrate the cube of NGC\,6720. \textit{Right}: Example of a
    fit realized on one of the integrated sky spectra of NGC\,6720
    that have been used to reconstruct the velocity calibration map
    shown in the left panel. Some diffuse emission of the interstellar
    medium surrounding the nebula can be seen in the spectrum. The
    position of these lines, which have been fitted with a different
    velocity parameter, is indicated. See text for details.}
    \label{fig:skymap-fig}
\end{figure*}

The only uncertainty in terms of velocity calibration of Fourier
Transform spectra is the zero point. The velocity zero point can be
estimated at each position in the field of view by observing a laser
source with a known wavelength. SITELLE's wavelength calibration is
based on a high-resolution laser cube taken with the telescope
pointing at zenith. As most of the observations are made with a
different gravity vector, the whole optical structure is subject to a
certain amount of distortion which changes the pixel-to-pixel zero
point. Also, the wavelength of the laser is expected to have a
long-term drift. A more refined wavelength calibration must therefore
be made by fitting the numerous Meinel OH bands present in the filter
passband. A velocity calibration map can then be reconstructed from
the OH lines velocity measured on integrated spectra that are taken
all over the field. The whole process has been implemented in ORCS
(Martin et al. 2016, in preparation). The sky velocity map used to
calibrate NGC\,6720 and an example of a fit realized on one of the
integrated sky spectra are shown in Fig.~\ref{fig:skymap-fig}. The
uncertainty on sky lines velocity in the integrated spectra is around
1\,\kms{}. An overall uncertainty on the pixel-to-pixel calibration of
0.5\,\kms{} appears to be a conservative estimation.

\section{Distribution of the posterior probability for a fit on a spectrum of NGC\,6720}

\begin{figure*}
  \includegraphics[width=\linewidth]{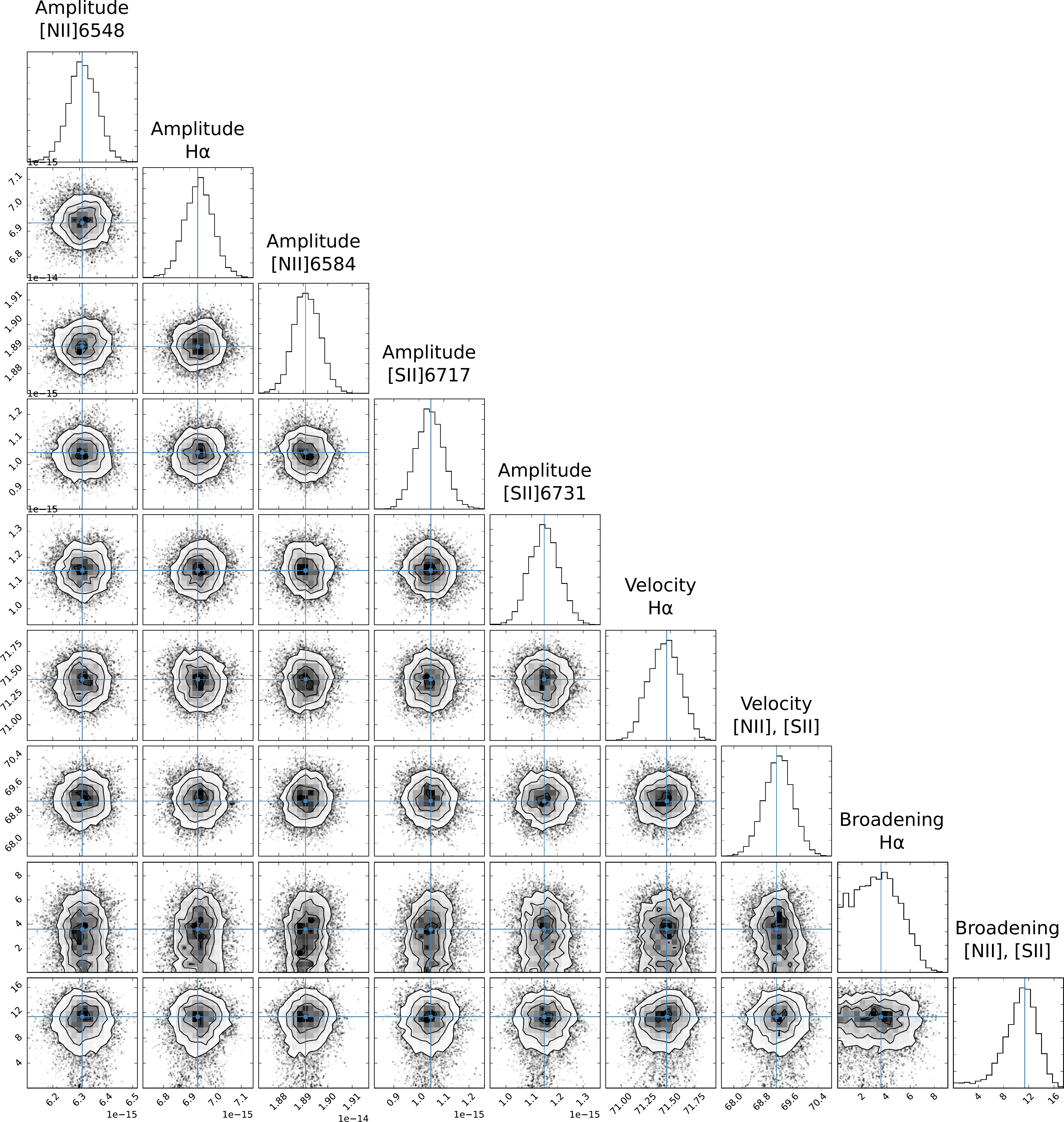}
  \caption{Corner plot of the marginalized distribution of the
    posterior probability for each parameter of a fit realized on the
    spectrum shown in Fig.~\ref{fig:m57_fit} with a Markov-Chain
    Monte-Carlo algorithm. See text for details.}
  \label{fig:triangle}
\end{figure*}

Fig.~\ref{fig:triangle} shows the marginalized distribution of the
posterior probability for each parameter of a fit realized on the
spectrum shown in Fig.~\ref{fig:m57_fit} with a Markov-Chain
Monte-Carlo algorithm. The fitted model is the sum of 5 lines
(H$\alpha$, [\NII]$\lambda$6548, [\NII]$\lambda$6584,
[\SII]$\lambda$6717, [\SII]$\lambda$6731) plus a constant continuum
(not shown here). Line model is the convolution of a sinc line and a
Gaussian broadening as described in Section~\ref{sec:ils}. [\NII] and
[\SII] lines are sharing the same velocity parameter and the same
broadening parameter while the H$\alpha$ line as its own velocity and
broadening. Amplitudes are fitted independently. FWHM is a fixed
parameter which depends only on the number of steps, i.e. on the
resolution (see Section~\ref{sec:ilssinc}). The diagonal of this
corner plot shows the marginalized distribution of the posterior
probability for each parameter independently. The other panels shows
the two-dimensional marginalized distribution for all couple of
parameters. This corner plot has been realized with the Python library
corner.py \citep{Foreman2016}.

\section{Uncertainty on the data}
\begin{figure*}
  \includegraphics[width=\linewidth]{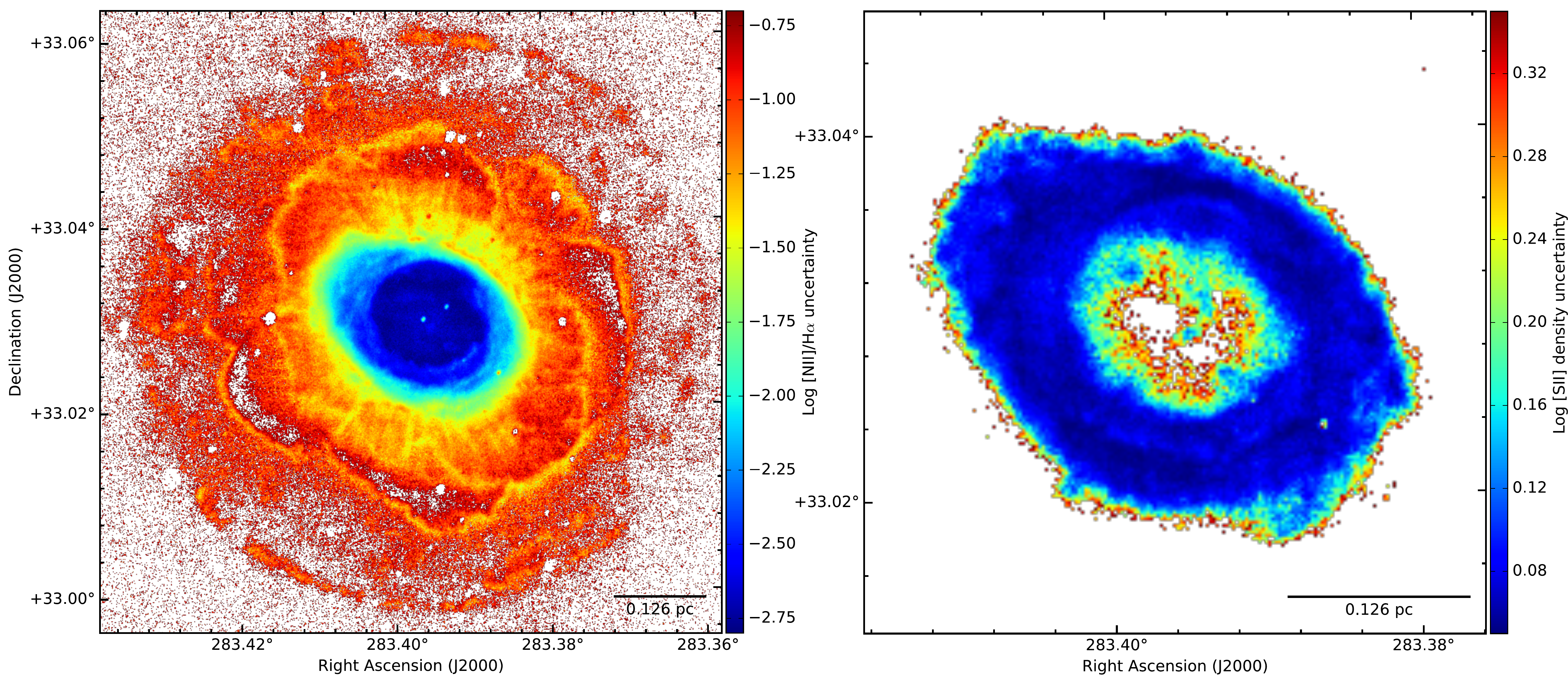}
  \caption{Uncertainty of the maps shown in Fig.~\ref{fig:ratios}. The
    order of the panels is the same. See text for details.}
  \label{fig:ratios-err}
\end{figure*}

\begin{figure*}
  \includegraphics[width=\linewidth]{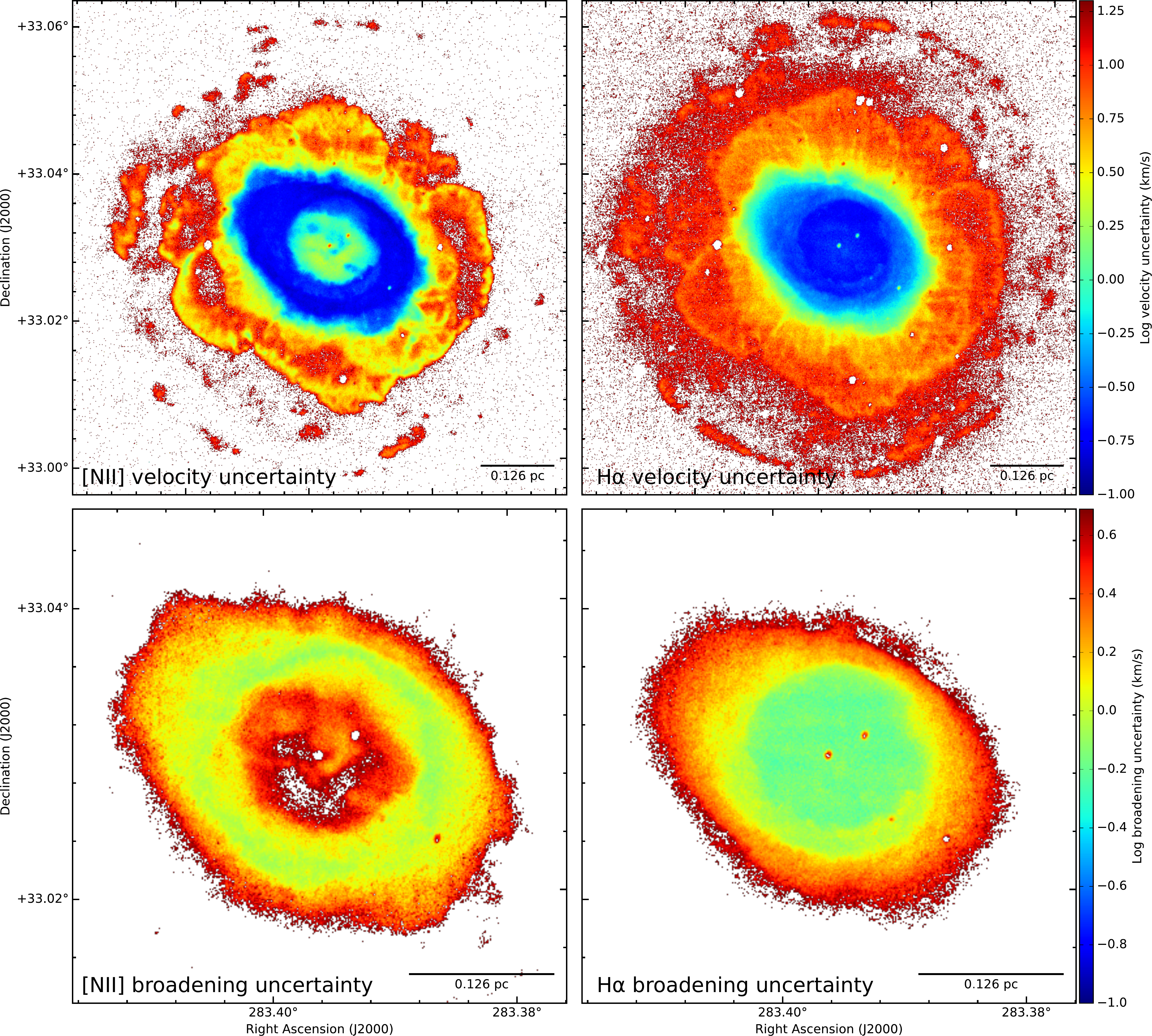}
  \caption{Uncertainty of the maps shown in Fig.~\ref{fig:maps}. The
    order of the panels is the same. See text for details.}
  \label{fig:maps-err}
\end{figure*}

Fig.~\ref{fig:ratios-err} and~\ref{fig:maps-err} show the uncertainty
on the maps shown respectively in Fig.~\ref{fig:ratios}
and~\ref{fig:maps}. The order of the panels is the same. The
uncertainty is calculated from the diagonal of the covariance matrix
returned by the function \texttt{scipy.optimize.curve\_fit} which
implements the Levenberg-Marquardt minimization algorithm
\citep{Jones2001}. 



\bsp	
\label{lastpage}
\end{document}